\documentclass[a4paper,fleqn,usenatbib]{mnras}
\usepackage[T1]{fontenc}
\usepackage{ae,aecompl}
\usepackage{graphicx}	
\usepackage{amsmath}	
\usepackage{amssymb}	
\usepackage{pdflscape}	
\usepackage{color}
\usepackage{ulem}
\usepackage{rotating}

\newcommand{\stkout}[1]{\ifmmode\text{\commout{\ensuremath{#1}}}\else\commout{#1}\fi}

\title[Kinematics and PLZ relation of RRLs]{Kinematics and Multi Band Period-Luminosity-Metallicity Relation of RR Lyrae Stars via Statistical Parallax}

\author[T. Dagne et al.]{
{Tesfaye Dagne Muhie,$^{1,2}$\thanks{E-mail: tesfayedagne7@gmail.com or tesfayed@bdu.edu.et}
A. K. Dambis,$^{3}$
L.N.Berdnikov,$^{3}$,
A.Y.Kniazev$^{4,5,3}$}
 and Eva~K.~ Grebel$^{6}$
\\
$^{1}$Astronomy and Astrophysics Research Division, Entoto Observatory and Research centre (EORC),\\ Ethiopian Space Science and Technology Institute (ESSTI), 
 P.O.Box 33679, Addis Ababa, Ethiopia\\
$^{2}$ Washera Geospace and Radar Science Research Laboratory, Department of Physics, Science College,\\  Bahir Dar University,
P.O.Box 79, Bahir Dar, Ethiopia\\
$^{3}$Sternberg Astronomical Institute, Lomonosov Moscow State University, Universitetskii pr. 13, Moscow,
119992 Russia\\
 $^{4}$South African Astronomical Observatory, PO Box 9, 7935 Observatory, Cape Town,South Africa,\\
$^{5}$Southern African Large Telescope Foundation, PO Box 9, 7935 Observatory, Cape Town, South Africa.\\
 $^6$ Zentrum f\"ur Astronomie der Universit\"at Heidelberg, Astronomisches Recheninstitut, M\"onchhofstr. 12-14, 69120, Heidelberg, Germany\\
}

\date{Accepted XXX. Received YYY; in original form ZZZ}

\pubyear{2021}

\begin{document}
\label{firstpage}
\pagerange{\pageref{firstpage}--\pageref{lastpage}}
\maketitle

\begin{abstract}
This paper presents results from photometric and statistical-parallax analysis of a sample of  850  field 
RR Lyrae (RRL) variables. The photometric and spectroscopic data for sample RRLs are obtained from (1) our new spectroscopic
 observations (for  448 RRLs) carried out with the Southern African Large Telescope (SALT);  (2)
 our photometric observations using  the 1.0-m telescope of the South African Astronomical Observatory (SAAO),  
 and  (3) literature.  These are combined with accurate proper motion data  from 
 the second release of \textit{Gaia} mission (DR2). This study primarily determines the velocity distribution of solar 
neighborhood RRLs, and it also calibrates  the zero points of the RRLs visual V-band luminosity-metallicity 
(LZ or $M_V-$[\text{Fe/H}]) relation and their period-luminosity-metallicity (PLZ) 
 relations  in  the \textit{WISE} $W_1-$ and \textit{2MASS} $Ks-$band. 
 We find the bulk velocity of the halo RRLs relative to the Sun to be   $(U_0,V_0,W_0)_{\text{Halo}}$ $=(-16\pm7,-219\pm7,-6\pm5)$\,km\,s$^{-1}$ in the direction of Galactic center, 
Galactic rotation and North Galactic pole, respectively,  with velocity-dispersion ellipsoids  $(\sigma V_R,
 \sigma V_\phi,\sigma V_\theta)_{\text{Halo}}=$ $(153\pm7,106\pm4,
 101\pm4)$\,km\,s$^{-1}$. The corresponding parameters for the disk component are found 
 to be $(U_0,V_0,W_0)_{\text{Disk}}=$ $(-19\pm5,-46 
\pm5,-14\pm3)$\,km\,s$^{-1}$ and $(\sigma V_R
 ,\sigma V_\phi,\sigma V_\theta)_{\text{Disk}}=$  
$(49\pm4,38\pm4,25\pm3 )$\,km\,s$^{-1}$. 
 The calibrated PLZ in $W_1-$, $Ks-$band, and V-band LZ relations  are   $\langle\,M_{W1}\rangle=$  $-0.824+0.124[Fe/H]-2.381\log\,P_F$, $\langle M_{Ks}\rangle=$ $-0.804+0.101[Fe/H]-2.33\log\,P_F$ 
and $\langle\,M_V\rangle=1.041+0.232[Fe/H]$, respectively.  The calibrated PLZ and LZ relations 
 are used to estimate the Galactic Center distance and  the distance
 modulus of the Large Magellanic Cloud (LMC), which are found to be  7.99$\pm$0.49\,kpc and 18.46$\pm$0.09 \,mag,   
 respectively. All our results
 are in excellent agreement with available literature  based on statistical parallax analysis, but are considerably more accurate and precise.  Moreover, the zero-points of our calibrated PLZ and LZ relations are quite consistent with current results found by other techniques and yield the LMC distance modulus that is within 0.04\,mag of the current most precise estimate.

\end{abstract}

\begin{keywords}
stars: distances -- stars: kinematics and dynamics -- stars: variables: RR Lyrae --Local Group -- distance scale.
\end{keywords}


\section{Introduction}
\label{sec:introd}
 RR Lyrae type variables are  low mass stars that have evolved off the main sequence when their core hydrogen is exhausted and  that
occupy a fairly narrow strip in the Hertzsprung-Russell diagram at the intersection of the pulsating-star instability
strip and the horizontal branch when their mean effective temperature reach $\sim6000-7500$ K \citep{iben1974post}. Thus, they are old low-mass yellow or white
giants in the core helium burning stage  \citep{salaris2005evolution} and belong to Population II \citep{smith1995rr,lipka2012rr}. Therefore,
they  can serve as distance indicators, as well as kinematics and metallicity tracers of old 
populations \citep[][and the reference there in]{dambis2013rr}. They are radially pulsating A--F  type stars which pulsate with periods 0.2-1.2 d \citep{catelan2009horizontal}. They were 
once stars with similar or slightly lower mass than the Sun, around 0.8 solar masses \citep{cacciari2012rr}. 
Most of them are classified into three main types according to the mode in which they pulsate radially: 
fundamental-mode RRab stars, first-overtone RRc stars and double-mode RRd stars. They are easily identified
by their periods and characteristic light-curve shapes. Like other pulsating stars they have their own PLZ relation in   any band \citep{nemec1994period,catelan2004rr}.

RRLs can be used as  standard candles because of their LZ and PLZ relations. The near- or mid-infrared 
wavelength (hereafter NIR or MIR)  PLZ relations are more preferable than the visual $M _{V}-[Fe/H]$ relation, 
 because they are less  dependent on interstellar 
extinction, metallicity and evolution effects.  
 For example, reddening in the visual V-band is one order of magnitude 
grater than in the K-band \citep[$A_{K}=0.114A_{V}$]{cardelli1989relationship}, 
and it is roughly 15 times stronger than in the W1-band 
\citep[$A_{W1}=0.065A_{V}$]{madore2013preliminary}. Moreover, the mean magnitudes of RRLs in NIR- and MIR-bands can be estimated more accurately than their mean visual V-band magnitudes due to the smaller amplitude  of infrared light 
curves \citep{jones1996template,monson2017standard}. 
 Despite these useful features of PLZ and LZ relations, there is still no agreement about their parameters especially concerning the zero points.
One of the problems is the persistent tension between the results coming from the statistical parallax methods with those from other techniques (pulsational models, Baade-Wesselnik, eclipsing binaries, RGB tip, etc.). We therefore try to recalibrate the zero points of visual V -band LZ and Ks- and W1-band PLZ relations of RRLs by applying the maximum likelihood version of statistical parallax technique to a sample with more than twice the size of the largest such list employed so far.

Statistical parallax is a primary method that can  determine simultaneously both 
 the distance scaling parameter (the mean absolute magnitude) and velocity 
ellipsoid parameters (like the bulk velocity and velocity dispersion) of a set 
of stars from observables including proper motions and radial velocities. 
Historically, statistical-parallax analysis of RRL stars was originally used by 
\cite{Pavlovskaya1953} and later was the subject of study by
many different authors \citep[e.g.][]{RigalJ1958,herk1965proper,hemenway1975absolute,
hawley1986absolute,
strugnell1986luminosity,layden1996absolute,tsujimoto1997absolute,fernley1998absolute,gould1998systematics, luri1998lmc,Popowski1998,clube1980statistical,kollmeier2013,
dambis2001,dambis2009,dambis2013rr,dambis2014mid,dambis2017rr}. It is difficult to show exactly how these different 
authors use the method independently, since they
share a common developmental history and employ similar kinematic modeling  procedures --- except a few earlier
studies all authors use what is known as the maximum
likelihood method. However, these studies do show clear differences in such details as the  
sample size used, the
numerical techniques used to maximize the likelihood function, the way in estimating the uncertainties of the
derived parameters, with some authors incorporating
additional features such as \citet{herk1965proper} 
using automatic rejection of outliers and \citet{luri1996new} considering the case of a multicomponent population \citep{layden1998stellar}.

\cite{Popowski1998} investigated how  the size of the sample,  observational errors,  
systematic errors  and other potential biases affect the statistical parallax solution by 
deriving analytic expressions for the uncertainties of each inferred parameter. 
According to their study, the error of statistical parallax solution parameters, like the 
distance scaling factor, generally behave as $\varpropto N^{-\frac{1}{2}} \kappa^{-1}$; where N is sample 
size, $\kappa=W/\sigma$ is the `Mach number' which depends on bulk velocity (W) and 
velocity dispersion ($\sigma $).  They also verified that the systematic errors for the solutions are generally small and these can be corrected using Monte Carlo simulations. Therefore, the accuracy of statistical parallax solution 
parameters improves as the sample of stars with accurate proper motions and radial 
velocities is increased in size. 

Hence building a large enough sample of RR Lyrae type variables with precise and homogeneous 
photometry, metallicity, proper motion and bona fide radial velocities is of prime 
importance for performing statistical parallax. Despite this fact, only about 400 plus RRLs 
in the extended solar neighborhood had all these data available \citep{dambis2013rr} until 
recently. To alleviate this we carried out the  MAGIC project \citep[see][for brief description 
of the project]{Kniazev2019}   with SALT \citep{buckley2006ground,o2006first} telescope
with the purpose of acquiring
the observationally expensive spectroscopic data for the greatest possible number of RRLs
in addition to our long-term program of photometric observations of these 
variables \citep{dambis2017rr}. Thus, our new radial velocity and metallicity data 
for 448 RRLs coming from this program as will be published by A.Y.Kniazev (in 
preparation) bring the size of the sample to  850 stars.

Furthermore, \textit{Gaia} DR2 has recently published multi-band photometry (\textit{Gaia} \textit{G}, $\textit{G}_{BP}$ and $\textit{G}_{RP}$ ) 
data as well as five-parameter astrometric (position, parallax, and proper motion) solutions for $\approx$1.3 billion sources 
as calculated solely from \textit{Gaia} astrometry  \citep{brown2018gaia}. For the sources with five-parameter \textit{Gaia} 
astrometric solutions the median uncertainty in parallax  at the reference epoch J2015.5
is about 0.04 mas for bright (\textit{G} < 14 mag) sources, 0.1 mas at \textit{G} = 17 mag, and 0.7 mas at \textit{G} = 20 mag, and 
the corresponding uncertainties in the proper motion components are 0.05, 0.2, and 1.2 mas $\text{yr}^{-1}$, 
respectively  \citep{lindegren2018gaia}. Furthermore, \textit{Gaia} proper motion data are more uniform and accurate 
than previously published catalogs. Therefore, it is now an opportune time for a new statistical parallax 
analysis in order to use the benefit of our more extensive spectroscopic
 data combined with  high quality \textit{Gaia} proper motion data.

In this study we estimate the velocity distribution 
 of Galactic field RRLs in the solar neighborhood, 
and refine the zero point of  the $M_V-[Fe/H]$ and PLZ relations in the \textit{WISE} \textit{W}1- and \textit{2MASS} \textit{K}s-bands 
via statistical parallax technique. We also use multiband photometric data and   estimates obtained from a  fide
3D interstellar extinction model \citep{drimmel2003three} to calibrate RRL $ (V-Ks)_0 $ and $(V-W_1)_0$ intrinsic colours in terms of fundamental period and metallicity by following the footsteps of  \cite{dambis2013rr}. The improvement in this 
study includes addition of our own new photometric and spectroscopic data for about 448 RRLs, which is 111 \% increase over the largest such sample studied so far combined with accurate \textit{Gaia} proper motion data.

This paper is organized as follows. Section~\ref{sec:Data} describes in detail the data used in our statistical parallax 
analysis and how it is compiled. Section~\ref{sec:Method} reminds the basic principles and provides an overview of statistical 
parallax technique. Section~\ref{sec:Results} provides  the main results of this work: the kinematic model of Galactic field RRLs; 
calibrated LZ and PLZ relations of RRLs, and discusses the implication of our results to determine \textit{Gaia} DR2 parallax 
 offset, Galactic center distance and LMC distance modulus. Finally, we summarize our conclusions in Section ~\ref{sec:Conclusions}.

\section{Data}
\label{sec:Data} 
 As we stated in our previous section, in order to determine
the kinematic model parameters and calibrate the $M_V-[Fe/H]$ and PLZ relations in the $Ks$- and $ W_1$-bands  using statistical 
parallax method with better precision, we need a large sample of RRLs with accurate and precise  (1) apparent mean 
magnitude in the Johnson V, \textit{2MASS}
Ks and \textit{WISE} $ W_1$ pass-bands;  (2) known
periods and pulsation modes;  (3) metallicities, proper motions, radial velocities and extinctions in
one of the pass-bands. Initially, our sample for this work consists of 850 RRLs. 
Depending on the spectroscopic data source, we subdivide this sample of RRLs into two parts: The first one is the most complete sample of Galactic field RRL type 
variables employed in our previous statistical parallax analysis  
\citep{dambis2013rr},  this list initially includes a total of 403 Galactic field RRL stars  but one RRL is discarded 
due to current updates (see Sect~\ref{sec:period}); the 
second part consists of 448 RRL type variables, which have our own new  spectroscopic data acquired  
 as a result of observations performed with SALT  telescope. 
The next subsections describe each type of data 
 in detail.

\subsection{Periods and pulsation modes}
\label{sec:period}
The accuracy of the phased data relies on the accuracy of the pulsation  periods. Thus, the periods and pulsation modes 
of the first part of the    RRLs samples used in this work were adopted from our previous study \citep{dambis2013rr},  at that 
time we took these pulsation periods from ASAS3 catalogue \citep{pojmanski2002all}, \cite{maintz2005proper} compilation 
and General Catalogue of Variable Stars \citep[GCVS,][]{Samus2012}. However, recently GCVS \citep{Samus2017}  updated 
parameter values for some RRLs in this sample. Following these updates,
we adopt different period values than in our previous study  for 28 RRLs (in the range from the longest SS Gru to 
shortest DI Aps periods), and the pulsation mode of
DH Peg and EZ Cep were changed to RRC and RRAB, respectively. 
Moreover, BI Tel (P=1.17 days) turned out to be  an Anomalous Cepheid  and was
hence discarded \citep{muraveva2018rr}. Our final sample from this group consists of 402 RRLs, of which 367 pulsate in 
the fundamental mode and 35 in the first overtone mode.

 Given that some RRLs show short time scale random period fluctuations, it is unsafe to use outdated published ephemerides
to calculate the phases of our SALT spectroscopic observations \citep[e.g.][]{berdnikov2017short}. We therefore determined the periods and pulsation modes for the stars of the
second  part of our RRL sample  
from the time 
series analysis of our own photometric observations 
obtained with  the 1-m telescopes of SAAO in 2016 as well as the data from 
recently published All-Sky Automated Survey for Supernovae (ASAS-SN) catalogue 
\citep{kochanek2017all}  with epochs close to our SALT  spectroscopic observations (see the next section for details).

\begin{figure}
\includegraphics[width=\columnwidth]{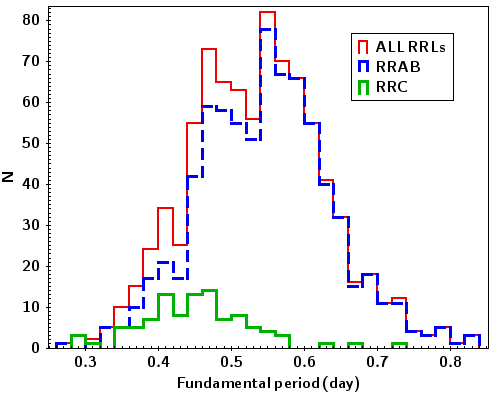}
    \caption{The distribution of fundamental periods.}
    \label{fig:Period_histog}
\end{figure}

The fundamental-mode periods ($P_{F}$) for first-overtone 
pulsators (99 RRc type variables) of our total sample  were computed as $\log P_{F}= \log P+0.127$ \citep{iben1974post}.
Figure~\ref{fig:Period_histog} shows the distribution of fundamental periods for our  total sample, with
the shortest period being P=0.25 days (DI Aps) and the
longest P=0.96 days (SS Gru). Thus our sample spans a representative range of RRL
periods and contains both RRab and RRc type stars.

\subsection{Apparent intensity-mean magnitudes in the V-, Ks- and W1-bands}
\label{sec:mean_mag}
 Like the case of the period, we  initially adopt the apparent intensity-mean magnitudes for the  first group of 402 RRL samples 
in the \textit{Johnson V}, \textit{2MASS} $K_s $ and \textit{WISE W}1 passbands from \cite{dambis2013rr} publication, and then update some of these values based on the resent literatures.
 According to that paper, (1) the \textit{V}-band intensity-averaged magnitudes were calculated from nine 
overlapping sets of observations \citep[see][and references therein for details]{dambis2013rr}; (2) the \textit{Ks}-band 
intensity-averaged magnitudes for most of stars were estimated by applying \cite{feast2008luminosities} phase correction 
procedure on their single-epoch Ks measurements of \textit{2MASS}  \citep{cutri20032mass}; (3)  the single-epoch \textit{2MASS}
Ks magnitudes without phase-corrections were adopted for 32 RRLs, which did not have ephemeris or it was too outdated; (4) the \textit{Ks}-band intensity-averaged magnitude 
for thirty stars were adopted from \cite{kinman2007,kinman2012}, and (5) the \textit{W1}-band intensity-averaged magnitudes   
for 398 RRL samples were computed (via Fourier fit) from the \textit{WISE} All-Sky data release \citep{cutri2012wise}. 

For the sake of clarity, the compilation of multi-wavelength (V, Ks, W1) mean magnitudes for our total
sample using our observations and archives data  are presented in the following subsections as organized by passband. 
Optical-, MIR- and  NIR-band data are described in
Subsection~\ref{sec:v_band}, Subsection~\ref{sec:w1_mag} and Subsection~\ref{sec:ks_mag}, respectively.

\subsubsection{V-band data}
\label{sec:v_band}
As we already pointed out, in order to complement our spectroscopic 
observations  for the second part of our sample, we  analyzed the photometry data obtained from our observations 
and  \textit{ASAS-SN} in order to determine the pulsation 
characteristics (period,
epoch of maximum light, amplitudes and  mean magnitudes)  of the corresponding RRLs.

\subsubsection*{Our photometric observations and Data Reduction}
\label{sec:observation}
We had performed photometric observations for a subset of  197 stars that belongs to the second  part of our sample
during two observing runs between March and May 2016 using 1-m telescope of SAAO equipped with $BVI_c$ filters of the 
Kron-Cousins system \citep{cousins1976}. During photometric nights we were observing two pairs of extinction stars 
or standard stars (one red and one blue: one pair near the zenith and the other near an air mass of two) in every 
two to three hours  interval \citep[according to the routine described in][and the references therein]{berdnikov2012}, 
in addition to visible target RRLs.  We use these observations  to calculate the extra atmospheric magnitudes 
of extinction stars by determining the atmospheric extinction coefficients.  We use the same measurements of these standard stars to derive the transformation coefficients and zero points, which helps us to transform the 
instrumental extra atmospheric magnitudes to the standard
Kron-Cousins $BVI_c$ system.

 We therefore used the same reduction routines as  those employed 
in \cite{berdnikov2012} to initially reduce the
CCD frames taken on all photometric night only. These would let us to determine seasonal mean transformation 
coefficients which would then be applied to observations of the same standards on each night in order to find 
the zero points for each night. We thus compiled a catalog of positions and magnitudes of all objects on the 
best CCD frames from
the reduction for all photometric nights.  We then selected constant stars from this catalog and used them as 
comparison stars for differential photometry of all
stars on all CCD frames including those taken on non-photometric nights. All this is done by using our own 
reduction package (custom software developed by L.Berdinkov).

\begin{figure}
\includegraphics[width=\columnwidth]{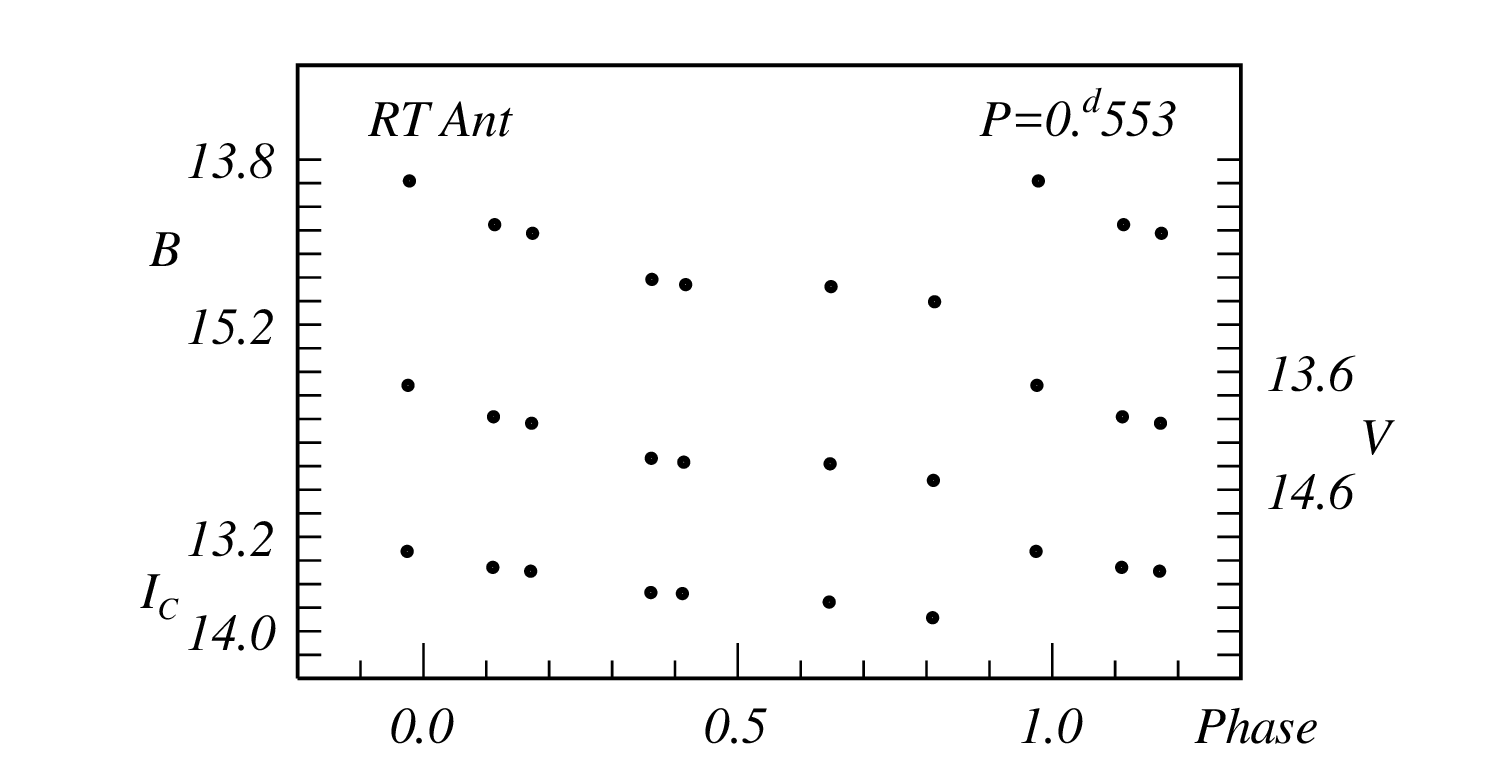}
    \caption{Phased light curve of RT Ant.}
    \label{fig:RT_ANT_LC}
\end{figure}

Our BV$I_c$ measurements for a typical example RT Ant are presented in Table~\ref{tab:RT_ANT}, and 
 the corresponding phased 
light curve is shown in Fig.~\ref{fig:RT_ANT_LC}. 
The accuracy of our individual observations is nearly 0.01 mag in all filters. Our measured data for 197 RRLs 
subsample (as illustrated in Fig.~\ref{fig:RT_ANT_LC}) are generally not sufficient to model their light curves 
accurately.  For this reason we complemented the time series analysis of our measured V-band  observations  with  \textit{ASAS-SN} data and applied the
 Lomb-Scargle Algorithm  \citep{barning1963numerical,lomb1976least,scargle1982studies}  to determine the initial 
ephemerides.  We then used these ephemerides  
to determine the amplitudes, intensity mean magnitudes (with uncertainties in the order of $ \pm0.01$ mag) and 
epochs of maximum light via the \cite{hertzsprung1919bearbeitung}  method employing the algorithm proposed by \cite{berdnikov1992study}.   
Additional light curve data and plots for 33 and  163 of  the samples using our observations had been published in \cite{dagne2017oc}  and  \cite{dagne2020}, respectively (available in their online version).

\begin{table*}
	\centering
	\caption{Photometric Observations of RT Ant}
	\label{tab:RT_ANT}
	\begin{tabular}{lcccccccr} 
		\hline
		 HJD 2400000 & Filter & Magnitude & HJD 2400000 & Filter & Magnitude & HJD 2400000 & Filter & Magnitude \\                                                                                                                                                                                                                                                                                                                                                                                                                                           
        \hline                                                                                                                                                                                                                                                                                                                                                                                                                                                                                                              
 57467.4702& $I_c$   &    13.885 &      57467.4708& $V$      &    14.521 & 57467.4719& $B$  &    15.006 \\
 57469.4620& $I_c$   &    13.680 &      57469.4632& $V$      &    14.366 & 57469.4648& $B$  &    14.860 \\
 57470.4345& $I_c$   &    13.491 &      57470.4353& $V$      &    14.036 & 57470.4361& $B$  &    14.425 \\
 57473.4618& $I_c$   &    13.752 &      57473.4625& $V$      &    14.380 & 57473.4634& $B$  &    14.877 \\
 57476.4837& $I_c$   &    13.458 &      57476.4843& $V$      &    13.981 & 57476.4854& $B$  &    14.352 \\
 57479.3875& $I_c$   &    13.672 &      57479.3878& $V$      &    14.333 & 57479.3884& $B$  &    14.816 \\
 57480.2791& $I_c$   &    13.323 &      57480.2799& $V$      &    13.715 & 57480.2812& $B$  &    13.981 \\
\hline                                                                                                                                                                                                                                                                                                                                                                                                                                                                                                              
	\end{tabular}
\end{table*}
 
\subsubsection*{ASAS-SN}
\label{sec:ASAS}
The \textit{ASAS-SN} project is working towards imaging the entire visible sky every 
night to a depth of V $\sim$ 17 mag; its aperture photometry data with $\sim100-400$ 
epochs of observation that spans $\sim 2-5$ years up to now are freely accessible at 
\textit{ASAS-SN} website \footnote{\url{https://asas-sn.osu.edu}}\citep{kochanek2017all}. 
Thus, compared to other surveys which have photometry from one to more than ten
years older than our SALT spectra measurement, \textit{ASAS-SN} data 
can give us pulsation parameters near the dates of our SALT spectroscopic measurements. 
 We therefore used the \textit{ASAS-SN} V-band data  with epochs
close enough to our SALT spectroscopy measurements  to determine the phases of these measurements. 

\begin{figure}
\includegraphics[width=\columnwidth]{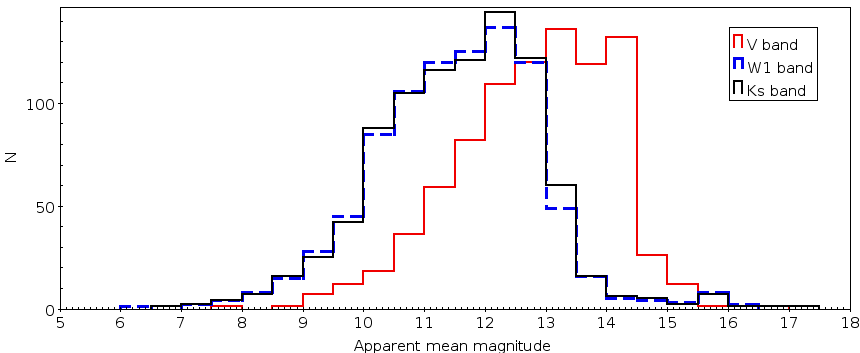}
    \caption{Apparent magnitudes distribution.}
    \label{fig:mean_V}
\end{figure}

 Note, however, that ASAS-SN data, which proved to be invaluable for determining the accurate phases of spectroscopic observations,
may yield not sufficiently accurate intensity-mean magnitudes because of the large pixel size ($\sim$~8~arcsec). On the other hand,
Gaia DR2 provides photometry in three bands ($G$, $B_P$, and $R_P$) based on about one dozen-plus visits. We used the homogenized $V$-band 
intensity-mean magnitudes from~\citep{dambis2013rr} combined with the intensity-mean magnitudes derived from our new CCD observations 
to calibrate $<V>$ in terms of
the average $B_P$, and $R_P$ magnitudes provided in Gaia DR2:
\begin{equation}
<V>= B_P+(0.146~\pm~0.047)-(0.763~\pm~0.112)(B_P-R_P)
\label{eq:Gaia_V_calibration}
\end{equation}
$$
+(0.268~\pm~0.063)(B_P-R_P)^2
$$
with a scatter of 0.109.
The  scatter is substantially smaller than the 3-$\sigma$-clipped standard deviation of the difference between the intensity-mean magnitudes 
inferred from ASAS-SN data and the estimates derived via equation~(\ref{eq:Gaia_V_calibration}), which is equal to $\sim$~0.17, and we therefore
adopt the intensity-mean $V$-band magnitudes converted from Gaia DR2 photometry for 251 stars lacking in our earlier list and without our CCD 
observations (including 19 stars from \citep{dambis2013rr} with no V-band data).
 
  Finally, these  homogeneous V-band mean magnitude data  together with their corresponding errors 
and  source of references for the total RRL samples  are provided in column 6 and 7 of 
Table \ref{tab:table_1}, respectively; their distribution is shown in Fig.~\ref{fig:mean_V}. 
As seen from Fig.~\ref{fig:mean_V}, the mean V-band magnitudes of our total sample range from 7.7 to 16.8 mag,  and almost all of our sample RRLs are brighter than 14.5 mag

\subsubsection{WISE W1-band data}
\label{sec:w1_mag}
The \textit{WISE} all- sky photometric survey mapped the entire sky in four MIR bands W1, W2, W3 and W4 with 
the effective wavelengths of 3.368, 4.618, 12.082 and 22.194 $\mu$m, respectively \citep{wright2010wide}. 
The most comprehensive data of the first two \textit{WISE} survey data releases are available 
as \textit{ALLWISE} catalog \citep{cutri2013explanatory}. \textit{WISE} operations are currently 
continuing with a mission called the "near-Earth object WISE reactivation (\textit{NEOWISE-R}) mission". 

Fortunately, the W1-band intensity mean magnitudes for 231 of our second group RRL samples were computed 
by \cite{chen2018wide} using Fourier fitting techniques  of \textit{ALLWISE} and \textit{NEOWISE-R} data.
 Furthermore, \cite{gavrilchenko2014mid}  estimated the W1-band intensity mean magnitudes for another 121 RRLs 
of our second group samples by applying robust light-curve template fitting  to \textit{ALLWISE} data. 
Therefore, we adopted the W1-mean magnitudes for 352 RRLs of our second subsample from  these two sources.   However, we adopted the single-exposure \textit{ALLWISE} W1-band  data as the \textit{WISE} W1-band
 mean magnitudes for the remaining 96 RRLs; this doesn't affect our analysis because the amplitudes of MIR
light curves are small. In order to obtain the single-exposure \textit{ALLWISE} data, we simply cross-matched the remaining part of our second RRL  subsample with 
\textit{ALLWISE} catalog using 2-arcsec radius. Here, we also adopted the W1-mean magnitudes for 398 RRLs of our first subsample from \cite{dambis2013rr}  estimates as mentioned earlier.

Finally, these adopted W1-band intensity-mean magnitude and single exposure data with their corresponding  errors and  source of references for all of our sample RRLs except four stars (V957 Aql, V524 Oph, HM Aql and IN Psc) with no W1 band measurement values,  are  available in column  13 and 14  of Table \ref{tab:table_1}, respectively, and  their distribution is shown in Fig.~\ref{fig:mean_V}.

\subsubsection{Ks-band data}
\label{sec:ks_mag}
 We use the intensity-mean Ks-band magnitudes inferred from single-epoch photometry in Ks-band available from \textit{2MASS} catalog \citep{skrutskie2006} by
applying the \cite{feast2008luminosities}  phase correction procedure for most of the stars of our sample. For 144 stars from the first part of the sample 
we use the intensity-mean estimates adopted from the recent work of \cite{layden2019infrared} --- these are inferred by combining the
$K_s$-band magnitudes phase corrected by \cite{feast2008luminosities} with
the results of dedicated photometric observations by \cite{layden2019infrared} themselves, \cite{skillen1993}, \cite{fernley1993}, and unpublished photometry reported by
\cite{fernley1998absolute}. We derived the $K_s$-band intensity means for 205 stars of the first part of the sample by phase correcting the \textit{2MASS} magnitudes as described
by \cite{feast2008luminosities} in our earlier paper \citep{dambis2009} and adopted the phase-corrected magnitudes
from \citep{kinman2007} and \cite{kinman2012}
for 14 and 6 stars of the first part of the sample, respectively. We further phase-corrected the \textit{2MASS} $K_s$-band magnitudes for 262 stars of the second
part of the sample using the procedure described by \cite{feast2008luminosities}. This makes up for a total of 631 RRLs
(almost three quarters of the sample) with bona fide intensity-mean $K_s$-band
magnitudes. We left the \textit{2MASS} $K_s$-band magnitudes for the remaining 219 stars (32 stars in the first part and 187 stars in the second part of the sample) unchanged
since we could not properly apply the phase corrections to these stars because of the absence of properly timed ephemerides covering the \textit{2MASS} observing epochs.
Note that in our case we cannot determine the improvement of the scatter of the PL relation due  to the use of phase-corrected magnitudes directly, but we
can still estimate it from the scatter of phase correction, which for 369 RR Lyraes from the catalog of  
\cite{dambis2013rr} is equal to $\sigma_{phase-correction}$~0.09~mag,  and the intrinsic scatter of the PL relation, which is equal to about 
$\sigma_{intrinsic}$~0.05~mag \citep{braga2018rr}. Given that
the phase corrections to \textit{2MASS} $K_s$ magnitudes are totally  uncorrelated with the intrinsic deviations from the PL relation, we have
$\sigma_{uncorrected(PL)}^2$~=~$\sigma_{intrinsic}^2$~+~$\sigma_{phase-correction}^2$~$\sim$~0.10~mag, or about
twice the intrinsic scatter. Note, however, that when deriving all the color relations involving $K_s$-band magnitudes
below in this paper we used only stars with genuine intensity means or phase-corrected magnitudes, whereas uncorrected
\textit{2MASS} $K_s$-band magnitudes are provided only for the sake of completeness.\\Finally, these $Ks$ band mean magnitudes  and  single-epoch magnitudes with their corresponding errors and source of references for our total sample RRLs are available in column 11 and 12 of Table \ref{tab:table_1}, respectively, and their distribution is shown in Fig.~\ref{fig:mean_V}.

\subsection{Proper Motions and parallaxes}
\label{sec:Proper} 
The \textit{Gaia} mission \citep{brown2018gaia}  with its
legacy of astrometric and photometric data for more than 1.3
billion objects in the Milky Way and beyond brings a revolution in Galactic astrophysics. The only source of proper motions and parallaxes for this study is the recently released \textit{Gaia} DR2 catalogue, which is freely available through the \textit{Gaia}
archive website\footnote{\url{http://archives.esac.esa.int/gaia}}. We cross-matched our total sample of RRLs against the \textit{Gaia} DR2 catalogue using a cross-match radius of 2-arcsec to retrieve the DR2 proper motions, parallaxes and their errors for our total sample RRLs.

\begin{figure}
\includegraphics[width=\columnwidth]{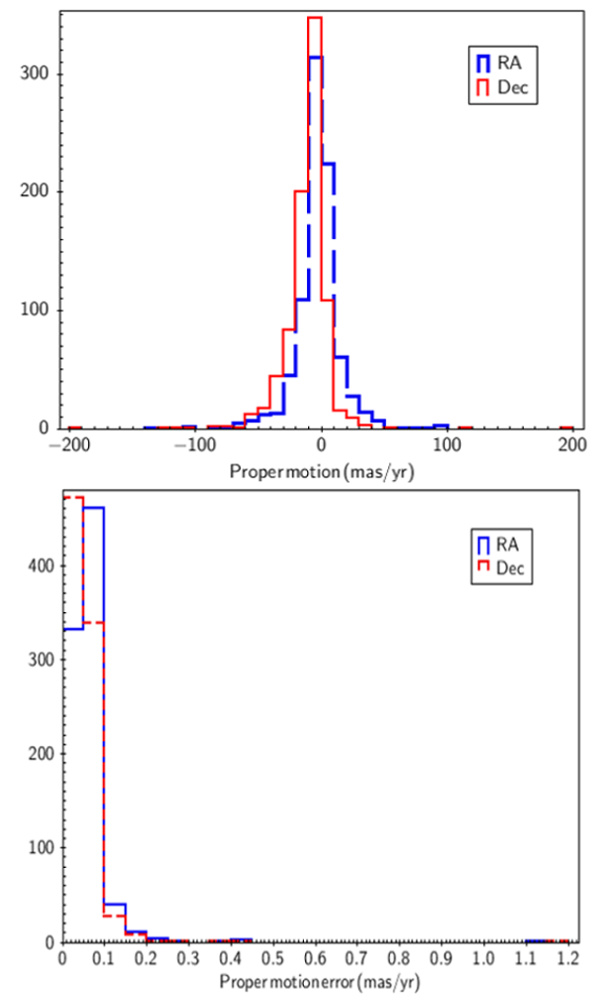}
    \caption{Distribution of proper motions and their error for our sample in both right ascension and  declination.}
    \label{fig:pm}
\end{figure}

\begin{figure}
\includegraphics[width=\columnwidth]{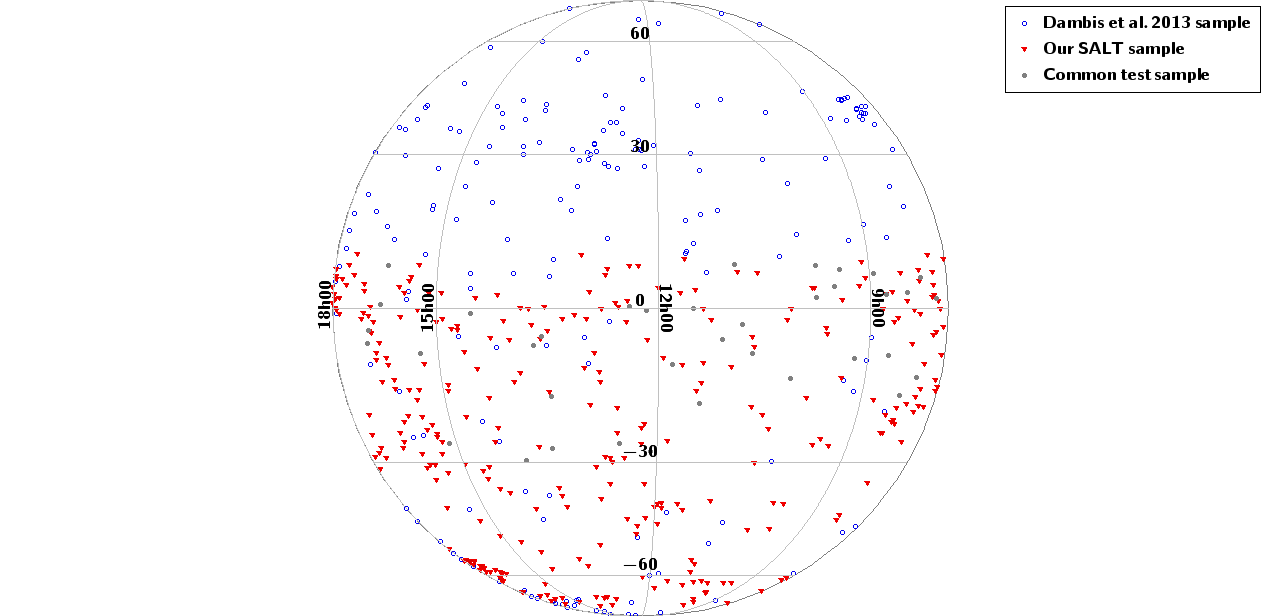}
    \caption{Sky distribution of the total 850 RRL samples. Our new addition samples from our SALT observation and Dambis et al., 2013 compilation are shown
with filled read triangles and open blue circles, respectively. 40 RRLs with both spectroscopic data which uses as the test sample for our observation are shown with filled gray circles.  }
    \label{fig:skydistribution}
\end{figure}

The distribution of the \textit{Gaia} DR2 proper motions and their uncertainties for our  entire RRL sample in right ascension and declination are shown in  Fig.~\ref{fig:pm}. We notice 
that the precision of proper motion for most of our sample RRLs ranges from 0.01
to 0.1 mas/yr and  smaller than 0.2 mas/yr for almost all stars. Thus, even the larger value of 0.2 mas/yr
 translates into a transversal velocity error of 5 kms$^{-1}$ at 5 kpc. 
Therefore, these accurate \textit{Gaia} DR2 proper motion data for our sample allow us to accurately determine their kinematics across their entire space even if we take into account of their errors. Moreover, the position of our sample RRLs as shown
in Fig.~\ref{fig:skydistribution} appear to be homogeneously distributed all over the sky; this  makes any possible spatially correlated systematic biases as small as  possible.

\subsection{Abundances}
\label{abund}
 Abundances data for our sample RRLs were not available in public release surveys like \textit{SDSS}, 
\textit{Gaia}  and others when we started this study.  Thus, like the case of other data types, we adopted homogenized metallicities on the 
\cite{zinn1984globular} scale from two primary sources: our previous compilation \citep{dambis2013rr} for the first part of our sample stars, and our new SALT  spectroscopic observations for the remaining  part. 
However, the metallicity for CK Uma  was not included in \cite{dambis2013rr} compilation. Therefore, we adopted the metallicity for this star from \textit{LAMOST}  survey \citep{luo2012data}. 

Our SALT  spectroscopic observations were performed 
using the Robert Stobie Spectrograph \citep[RSS;][]{burgh2003prime,kobulnicky2003prime} in the long-slit 
mode. The VPH grating GR1300 and 1.25 arcsec slit width were used for observations  
during 2016 and 2017. This configuration produces spectral resolution \mbox{R = 1300-1500}
and covers spectral range of 3900--6000~\AA , which gives us a possibility
to detect Balmer lines H$\alpha$, H$\beta$, H$\gamma$ and CaII\,K $\lambda$3924 line. These SALT spectroscopic observations were primary reduced using the SALT science pipeline \citep{crawford2010pysalt} as the long-slit reduction were done automatically in the way described in \citet{kniazev2008metallicity} with our pipeline based on  MIDAS, Unix Shell and IRAF programs. Here, we also used ULySS program by \cite{koleva2009ulyss} with a medium
spectral-resolution library of \cite{prugniel2011atmospheric} to determine
simultaneously the line-of-sight velocities, T$_{eff}$, log~g and [Fe/H] for  the observed stars.
Details of these reduction routines and their  accuracy with our RSS data for a particular MAGIG project star was presented
in \cite{Kniazev2019} and more detail for RRL samples of the project will also be presented in a forthcoming paper (Kniazev et al. 2020, in preparation).

\begin{figure}
\includegraphics[width=\columnwidth]{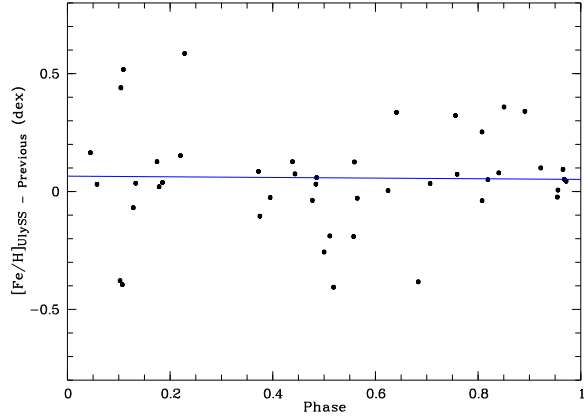}
    \caption{The difference between our abundance estimates based on SALT spectra and the previously published values for the calibrating subsample. Note the absence of any appreciable phase trend.  }
    \label{fig:feh_phase}
\end{figure}

In order to make the measurement scale of our SALT abundances homogeneous with our previous 
compilation scale, we calibrated our pipeline abundance measurements to our previous 
published \cite{zinn1984globular}
scale using subsample of RRLs, which have abundance measurement in both scales.  Based on 
this calibration, we find SALT metallicity scale  transformation equation~(\ref{eq:A1}) to 
\cite{zinn1984globular} metalliciity scale:
\begin{equation}
[Fe/H]_{ZW}=(0.836~\pm~0.053)[Fe/H]_{SALT}-(0.323~\pm~0.082)
\label{eq:A1}
\end{equation}
with a scatter of 0.23dex. Our abundances are measured at  random phase and we apply no phase corrections. The point is that although generally the inferred spectroscopic abundances 
of RR Lyraes vary with phase \citep{pancino2015chemical}, in our particular case the abundances inferred using ULySS program by \cite{koleva2009ulyss} combined 
with the medium of  \citet{prugniel2011atmospheric} appear to be phase independent (see Fig.~\ref{fig:feh_phase}, where the difference between our SALT RR Lyrae 
abundances and previous measurements is plotted against the phase of our SALT observations).

Whenever necessary, we also made transformation to the modern scale of \cite{carretta2009intrinsic} via equation
\begin{equation}
[Fe/H]_{Carr}=1.105[Fe/H]_{ZW}+0.160
\label{eq:A2}
\end{equation}
 with a slightly increased scatter of 0.25dex. The formal internal errors of the inferred metallicities are too small (generally on the order of 0.01--0.05) to be realistic, and we therefore estimate errors
based on a comparison with published data (see equation~\ref{eq:A1} above). We assume that the average errors of the adopted metallicities are the same for
all stars whether  based on 
SALT spectra or on published data as presented in our earlier list \citep{dambis2013rr}. Given the scatter of equation~\ref{eq:A1} (0.23dex), the adopted
metallicities are accurate to within 0.23dex/$\sqrt{2}$~=~0.16dex.

\begin{figure}
\includegraphics[width=\columnwidth]{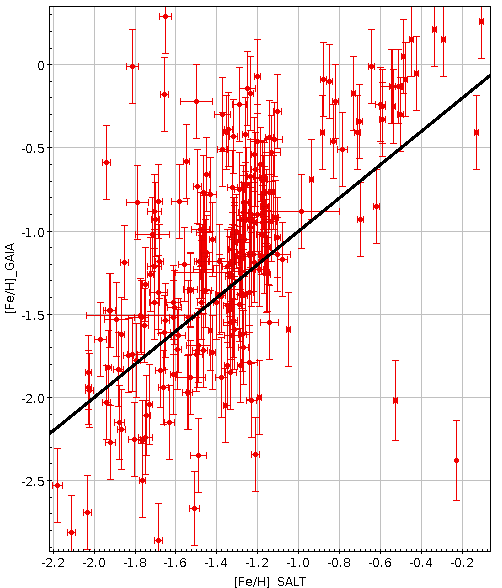}
    \caption{Comparison between our  SALT [Fe/H] with their available \textit{Gaia} DR2 data. The solid line represent the bisector}
    \label{fig:FE_SALT_VS_GAIA}
\end{figure}

Fig.~\ref{fig:FE_SALT_VS_GAIA} compares our  measured metallicities transformed
to the \cite{carretta2009intrinsic} scale (equations~\ref{eq:A1} and \ref{eq:A2} above)  with their available \textit{Gaia} DR2 metallicity \citep{clementini2019gaia}, which were estimated from the Fourier coefficients of \textit{Gaia} photometric data. Thus,  our metallicity measurements and \textit{Gaia} DR2 data generally agree for stars which have reliable \textit{Gaia} photometry, but our measurements are more  prices.

\begin{figure}
\includegraphics[width=\columnwidth]{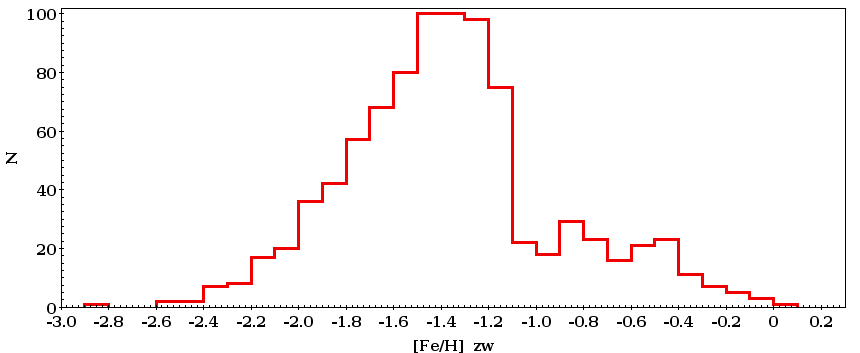}
    \caption{Distribution of [Fe/H] values for RRL variables of our total sample transformed to the  Zinn and West scale.}
    \label{fig:metalicity}
\end{figure}

Finally, these homogeneous metallicities  on
the \citet{zinn1984globular} scale with their corresponding errors and source of references for our total 850 sample  RRLs  
 are available in column 7 and 8 of Table  \ref{tab:table_1}, respectively. As can be seen in Fig.~\ref{fig:metalicity}, 
the metallicity distribution of the total sample span quite a broad range; namely, from  
$-$2.84 to +0.07 dex, and its peak lies between -1.2 and -1.5  dex.

\subsection{Radial velocities}
\label{radvelo}
Like for metallicity, we adopt the mean radial velocities $(V_R)$ and standard errors $(\sigma_{V_R})$ for the first and second  parts of our RRL sample from our previous 
publication \citep{dambis2013rr} and  from our SALT spectroscopic, respectively. We used our SALT H$\alpha$, H$\beta$, H$\gamma$ (Balmer) and CaIIK spectral line observations first to estimate the measured radial velocity at specific phases of 
pulsation. Then, the precise systematic (center of mass) velocities in the radial 
direction of RRLs are determined by subtracting the radial velocity due to pulsation at same phase from our highly precise measured radial velocities. 

\begin{figure}
\includegraphics[width=\columnwidth]{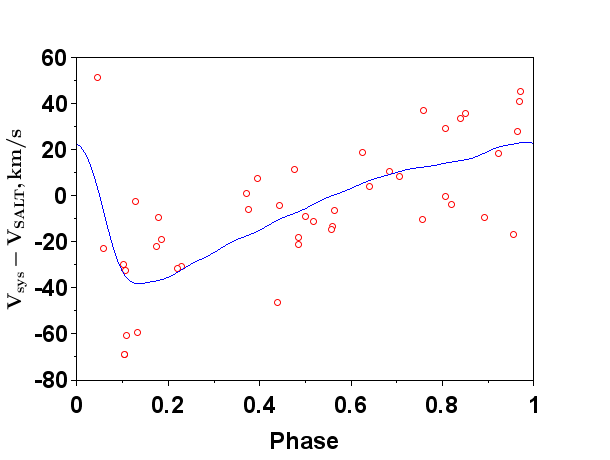}
\caption{Dependence of the velocity correction, $V_{sys}$~-~$V_{SALT}$, on   the phase of spectroscopic observations for the test sample stars.
The fit of the \citet{liu1991synthetic}  template scaled and shifted in both coordinates (see equation~\ref{eq:A3}) is shown by the continuous line.}
    \label{fig:Phase_delta_V}
\end{figure}

Given that the combination of spectral lines and the procedure we use to
measure the radial velocities differ from those employed by other authors, we cannot apply the published formulas and scaling procedures verbatim. The 
proper way would be to derive our own template radial-velocity curve and scaling relations by observing some stars at several phases with good enough sampling. 
Unfortunately, this approach was inapplicable in our case because most of the stars with SALT data (461 out of 488) were observed only once, 25 stars were observed 
twice, and one star was observed three times. We therefore decided to use a published template by properly scaling and shifting it in phase (applying a phase lag).
Fig.~\ref{fig:Phase_delta_V} shows the velocity corrections for the test sample (44  measurements for 40 stars from the list of \citet{dambis2013rr} with SALT observation) plotted against 
phase. A clear phase 
dependence is immediately apparent, which we fitted by the \cite{liu1991synthetic} template  $- S(\phi)-$ by finding the scaling factor and 
appropriate offsets in both 
phase and velocity difference. The least-squares fit yields the following result:

\begin{equation}
V_R{sys}-V_R(SALT) = -(38.6~\pm~5.8) + 
\label{eq:A3}
\end{equation}
$$
(62.0~\pm~9.3) S(\phi - 0.130~\pm~0.015)
$$
with a scatter of 20.5~km/s. 

\begin{figure}
\includegraphics[width=\columnwidth]{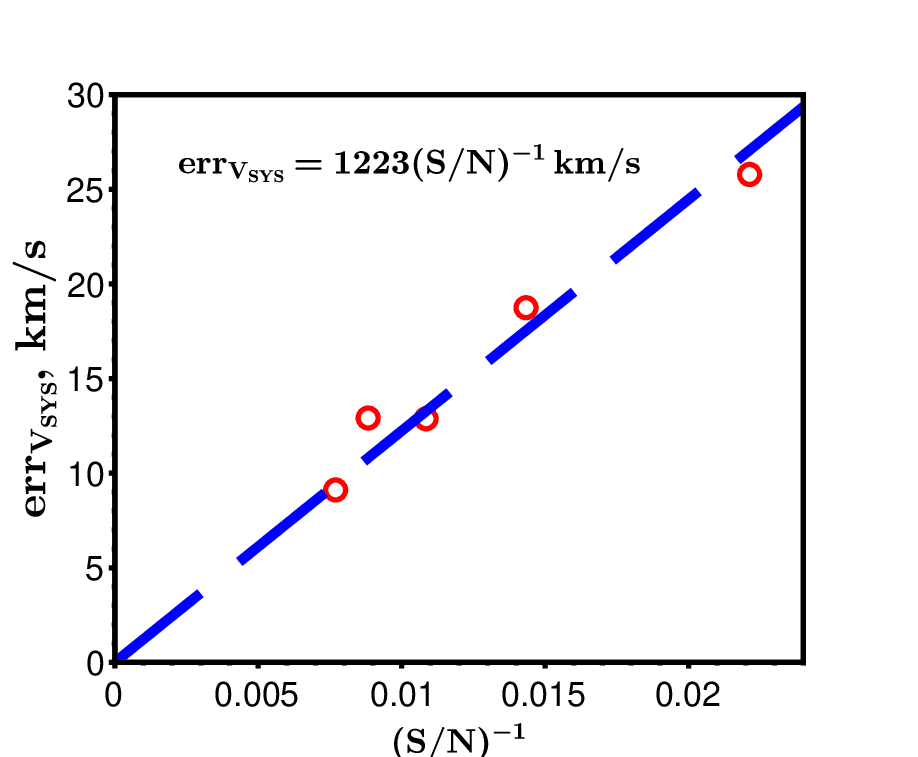}
\caption{ Dependence of the scatter of the previously published systemic velocity, $V_{sys}$, and our phase-corrected radial velocity, $V_{SALT}$, 
on signal-to-noise ratio.}
    \label{fig:err_vr_snr}
\end{figure}

 One would naturally expect the errors of $V_R{sys}$ inferred from $V_R(SALT)$ via equation~\ref{eq:A3} to depend on the quality of spectra
(quantified by the signal-to-noise ratio). We binned our test list of SALT radial-velocity measurements for RR Lyraes with published systemic velocities into five 
bins by signal-to-noise ratio and found the scatter of the difference between the published systemic velocities and those inferred from SALT data 
(after quadratically subtracting the errors of published velocities) to be inversely proportional to the signal-to-noise error, $\sigma_V(SALT)~\sim~1223/SNR$ 
(see Fig.~\ref{fig:err_vr_snr}). We use this relation to estimate the errors of all individual radial-velocity measurements  based on SALT data. These measurements have 
SNR values spanning from 11.4 to 154.4, implying a range of radial-velocity errors from 8 to 107~km/s. Because of the repeated measurements for a number of stars 
the errors of the mean radial-velocity estimates actually span from 5.6 to 107 km/s with an average value of 18.8 km/s.

\begin{figure}
\includegraphics[width=\columnwidth]{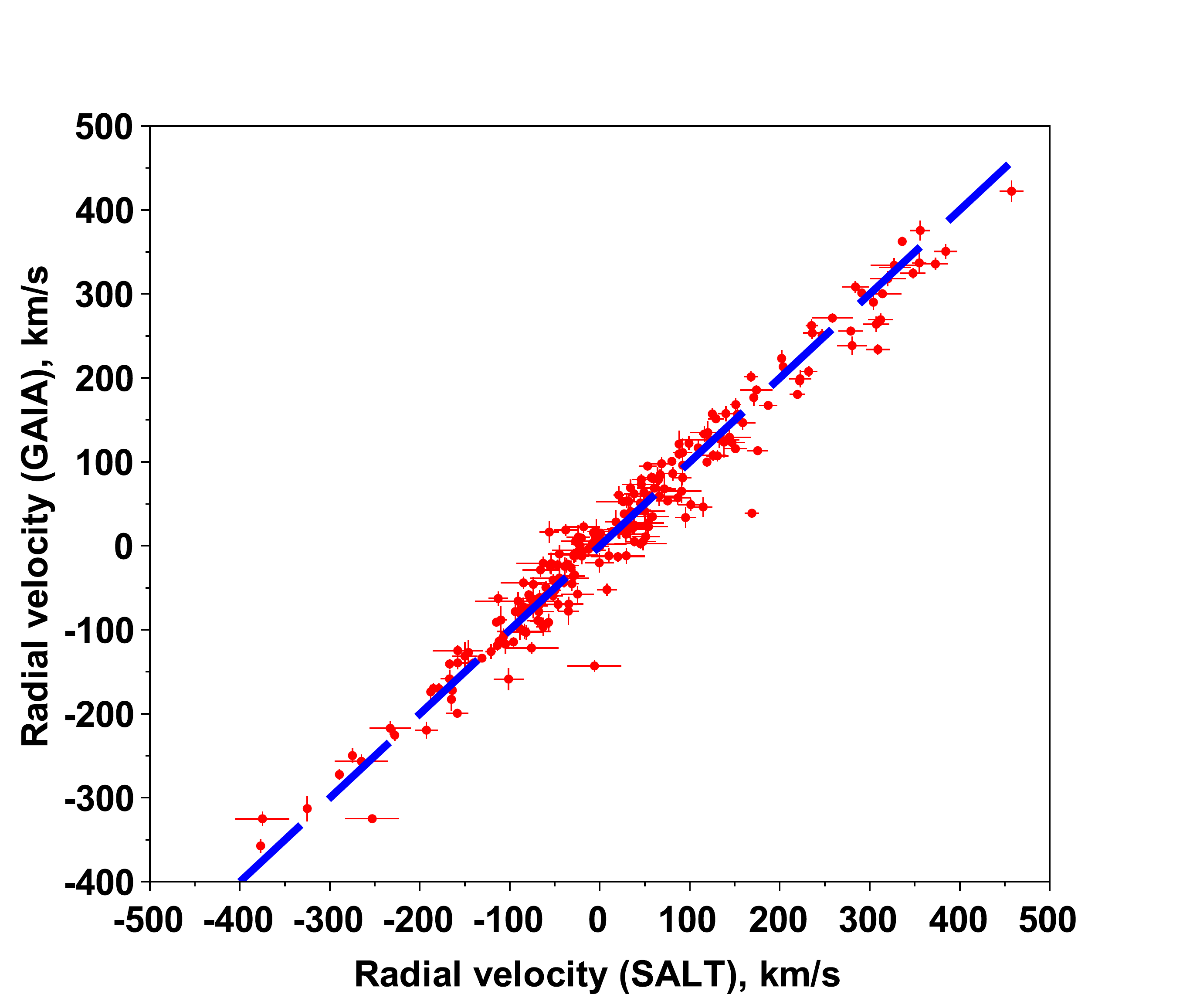}
    \caption{Comparison between our measured radial velocity  with their available \textit{Gaia} DR2 data.}
    \label{fig:V_R_SALT_VS_GAIA}
\end{figure}

Fig.~\ref{fig:V_R_SALT_VS_GAIA}  compares our measured radial velocities  with  
available \textit{Gaia} DR2 data  \citep{sartoretti2018gaia}. Thus our radial velocity measurements are in excellent agreement with available \textit{Gaia} DR2 data.

\begin{figure}
\includegraphics[width=\columnwidth]{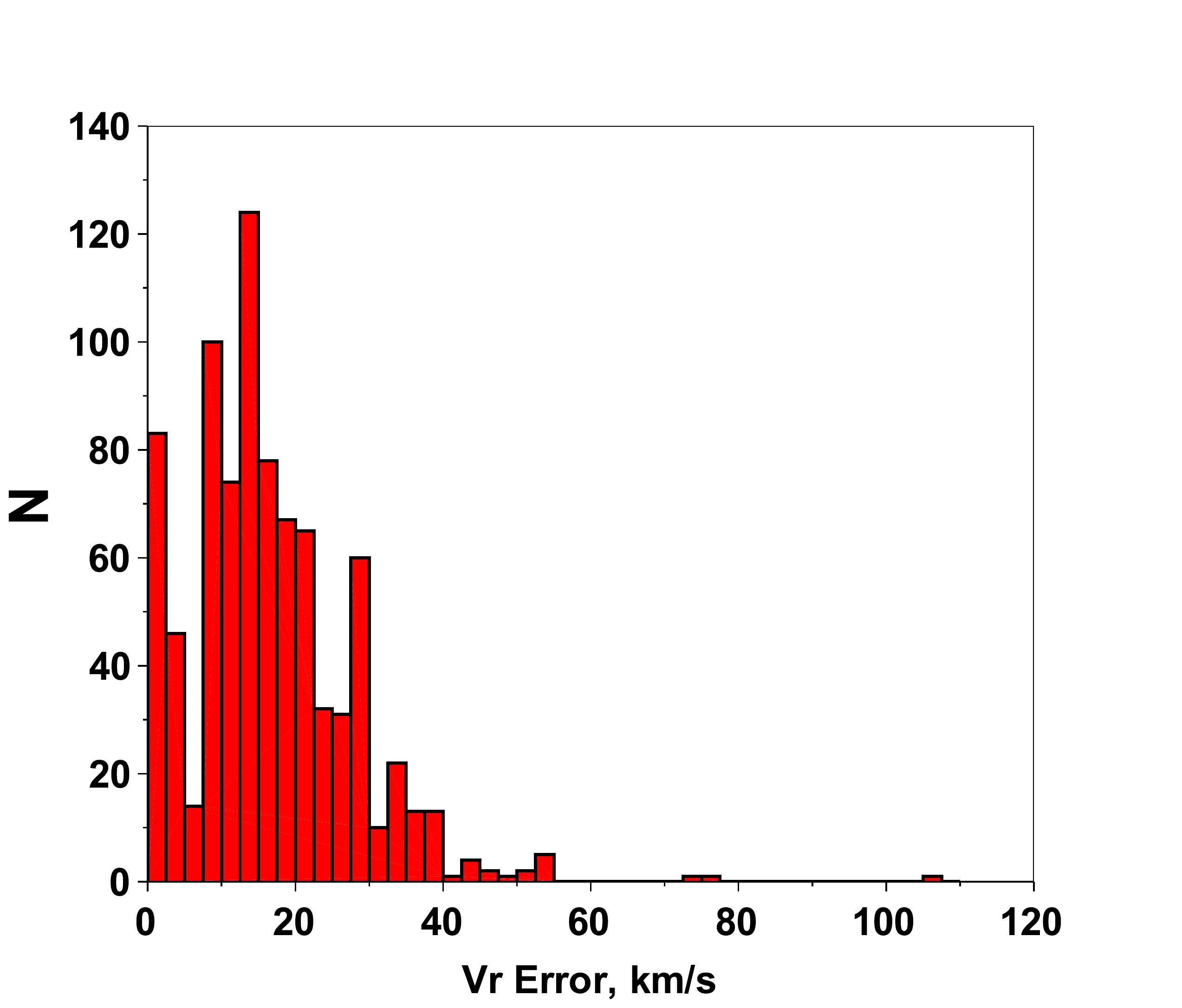}
    \caption{Distribution of radial velocity errors for sample stars.}
    \label{fig:error_Vr}
\end{figure}

The systemic (center of mass) radial velocities  $(V_R)$   with standard errors and the corresponding references for our total sample RRLs are presented in columns  15 and 16 of Table 
\ref{tab:table_1}, respectively. Fig.~\ref{fig:error_Vr} shows the distribution of radial 
velocity errors for all of our sample stars. As noted from Fig.~\ref{fig:error_Vr}, the  precision of radial velocities for most of our sample RRLs are better than  16km/s thereby contributing to accurate modeling of the kinematics of RRLs.

\subsection{Interstellar Extinction and Intrinsic colour calibrations}
\label{sec:extinction} 

Similarly to our previous work \citep{dambis2013rr},  we begin by estimating interstellar extinction toward our RRLs via the \cite{drimmel2003three} 3D extinction model using the iterative procedure described
in detail in that paper. The procedure consists of  (1) setting the starting extinction  values 
$A_V$~=~$A_{W1}$~=~$A_{KS}$
to zero; (2) computing the provisional photometric distances to 
the stars based on their apparent infrared magnitudes, the corresponding absolute magnitudes determined via 
the appropriate period-metallicity-luminosity relation,
and the currently adopted provisional extinction values; (3) updating  the $A_V$  extinction values 
by estimating them via the 3D map of \cite{drimmel2003three} based on the 
sky positions and provisional distances inferred at stage (2) and computing the $A_{W1}$ and $A_{Ks}$  
extinction values by converting the inferred $A_V$  using the \cite{yuan2013empirical} 
extinction coefficients $R_V=3.1$, $R_{W1}=0.18$, and $R_{Ks}$=0.306, and (4) repeating steps (2)--(3) with the 
updated extinction values until the latter 
converge. 
We adopt the \cite{dambis2014mid} WISE W1-band
PLZ relation in \cite{zinn1984globular} metallicity scale:
\begin{equation}
\begin{array}{rcl}
\langle M_{W1}\rangle &=& -0.814+0.106[Fe/H]-2.381 \log P_F, \label{eq:E1}
\end{array}
\end{equation}
 to compute the initial distances for the subsample of RRLs (846 stars) 
that have either measured intensity-mean magnitude or single exposure 
data in $W1$-band. For the four RRLs without the $W1$-band magnitude
data, we used the corresponding  period-metallicity-$M_Ks$ relation
\begin{equation}
\begin{array}{rcl}
\langle M_{Ks}\rangle &=& -0.769+0.088[Fe/H]-2.33 \log P_F, \label{eq:E2}
\end{array}
\end{equation}
derived by \citet{dambis2013rr}.

In principle, we could use the $A_V$, $A_{Ks}$, and $A_{W1}$ extinction estimates  so derived to calibrate the intrinsic colors
$(V-Ks)_0$ and $(V-W1)_0$ in terms of fundamentalized period and metallicity. 
The procedure is as follows. First, deredden the  intensity-mean magnitudes of our RR Lyraes in the W1-band $-\langle m_{W1}\rangle$, 
 Ks-band $-\langle m_{Ks}\rangle-$ and $V$-band $-\langle m_{V}\rangle-$ by their corresponding interstellar extinction values ($A_{W1}$, $A_{Ks}$  and $A_V$, respectively):
\begin{eqnarray}
\langle W1 \rangle_0&=&\langle m_{W1} \rangle-A_{W1} \label{eq:E5}\\
\langle Ks \rangle_0&=&\langle m_{Ks} \rangle-A_{Ks} \label{eq:E4}\\
\langle V\rangle_0&=&\langle m_V \rangle-A_V \label{eq:E3}
\end{eqnarray}
We then use equations{~(\ref{eq:E5}),~(\ref{eq:E4}) and~(\ref{eq:E3}) to compute} the intrinsic colours:
\begin{equation}
\begin{array}{rcl}
(V-Ks)_0 &=& \langle V \rangle_0-\langle Ks \rangle_0 \label{eq:E6}\\
 &=& \langle m_V \rangle-\langle m_{Ks} \rangle-(A_V-A_{Ks})
\end{array}
\end{equation}
\begin{equation}
\begin{array}{rcl}
(V-W1)_0 &=& \langle V \rangle_0-\langle W1 \rangle_0 \label{eq:E8}\\
 &=& \langle m_V \rangle-\langle m_{W1} \rangle-(A_V-A_{W1})  
\end{array}
\end{equation}
We now calibrate the $(V-W1)_0$ and $(V-Ks)_0$ colours  using the following
two well-established assumptions: first, that  $M_V$ depends linearly on $[Fe/H]$ and
does not depend on period at constant $[Fe/H]$ \cite[see, e.g.,][]{catelan2004rr}:
\begin{equation}
\langle M_V\rangle=a_V+b_V[Fe/H].
\label{eq:E9}
\end{equation}
and, second, that $<M_{W1}>$ and $<M_{Ks}>$ depend linearly on $[Fe/H]$ and log~$P_F$:
\begin{equation}
\langle M_{Ks}\rangle=a_K+b_K[Fe/H]+c_K\log P_F
\label{eq:E10}
\end{equation}
and
\begin{equation}
\langle M_{W1}\rangle=a_W+b_W[Fe/H]+c_W\log P_F
\label{eq:E11}
\end{equation}
 We further adopt the metallicity slope $b_V$~=~0.232 for the $V$-band metallicity-luminosity relation, so that:
\begin{equation}
\langle M_V\rangle=a_V+0.232[Fe/H].
\label{eq:E12}
\end{equation}
This slope is
 based on the two direct estimates \citep{gratton2004metal}, which involve  RR~Lyrae in the  LMC (believed to be at approximately
the same distance) and the the horizontal branches
in M31 satellite galaxies \citep{Federici2012} (also believed to be at approximately the same distance from us), respectively. As we pointed out earlier \citep{dambis2013rr},
the adopted metallicity slope is
also quite consistent with the estimates based on Baade-Wesselink analysis \citep{cacciari1992absolute}. 
We finally adopt the period slopes of the  2MASS $Ks$- and WISE $W1$-band period-metallicity-luminosity
relations from \cite{jones1992baade} and \cite{dambis2014mid}, respectively, where they are 
estimated from the data for RR~Lyraes in globular clusters, so that  equations~\ref{eq:E10} and \ref{eq:E11} become:
\begin{equation}
\langle M_{Ks}\rangle=a_K+b_K[Fe/H]-2.33\log P_F
\label{eq:E13}
\end{equation}
and
\begin{equation}
\langle M_{W1}\rangle=a_W+b_W[Fe/H]-2.381\log P_F,
\label{eq:E14}
\end{equation}
respectively.
We now subtract equations~(\ref{eq:E13}) and (\ref{eq:E14}) from equation~(\ref{eq:E12}) to obtain:
\begin{equation}
\begin{array}{rcl}
(V-Ks)_0 &=& \langle M_V \rangle-\langle M_{Ks} \rangle \label{eq:E15}\\
  &=& (a_V-a_K)+(0.232-b_K)[Fe/\\ 
  & &H] +2.33\log P_F.
\end{array}
\end{equation}
and 
\begin{equation}
\begin{array}{rcl}
(V-W1)_0 &=& \langle M_V \rangle-\langle M_{W1} \rangle \label{eq:E16}\\
  &=& (a_V-a_W)+(0.232-b_W)[Fe/\\ 
  & &H] +2.381\log P_F.
\end{array}
\end{equation}
We then move the period terms to the left-hand side:
\begin{equation}
\label{eq:E15a}
(V-Ks)_0-2.33\log P_F = (a_V-a_K)+(0.232-b_K)[Fe/H] .
\end{equation}
and 
\begin{equation}
\label{eq:E16a}
(V-W1)_0 -2.381\log P_F = (a_V-a_W)+(0.232-b_W)[Fe/H].
\end{equation}
We can now solve sets of linear equations~(\ref{eq:E15a}) and (\ref{eq:E16a}) with the unknowns being
$a_V-a_K$, (0.232-$b_K$)  and $a_V-a_W$,(0.232-$b_W$) for our RR Lyraes to obtain the
desired calibrations of 
the intrinsic colours $(V-Ks)_0$  and $(V-W1)_0$ in terms of fundamentalized period and metallicity. 
The resulting 3$\sigma$-clipped least-squares solutions are
\begin{equation}
\begin{array}{rcl}
(V-Ks)_0&=&1.751(\pm0.020)+0.115(\pm0.013)[\\
         & & Fe/H]+2.33\log P_F \label{eq:E15b}\\
\end{array}
\end{equation}
with a scatter of $\sigma(V-Ks)_0=0.175$ (791 of 850 stars)
and
\begin{equation}
\begin{array}{rcl}
(V-W1)_0&=&1.903(\pm0.016)+0.155(\pm0.011)[\\
         & & Fe/H]+2.381\log P_F \label{eq:E16b}\\
\end{array}
\end{equation}
with a scatter of $\sigma(V-W1)_0=0.137$ (768 of 846 stars).

\begin{figure}
\includegraphics[width=\columnwidth]{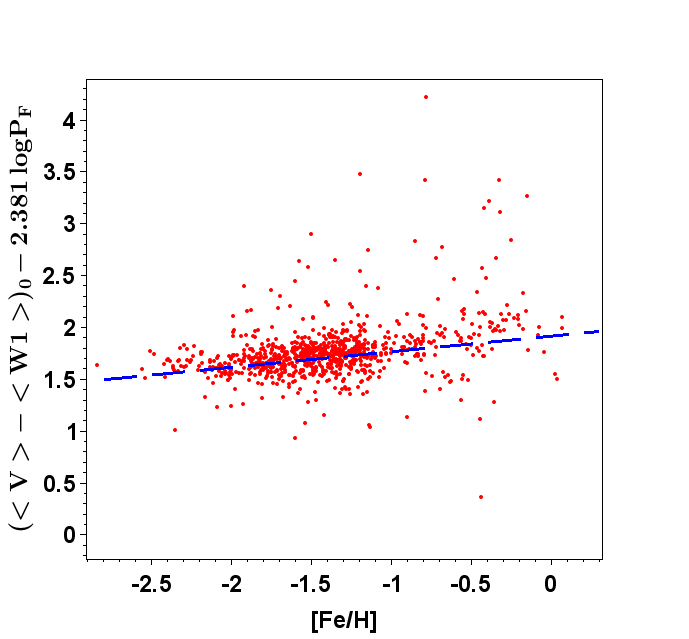}
    \caption{Plot of $(V-W1)_0-2.381\log P_F$ as a function of $[Fe/H] $. The
solid line shows the linear fit (equation~(\ref{eq:E16a})) with parameters~(\ref{eq:E16b})}
    \label{fig:V_W1_Fe}
\end{figure}
\begin{figure}
\includegraphics[width=\columnwidth]{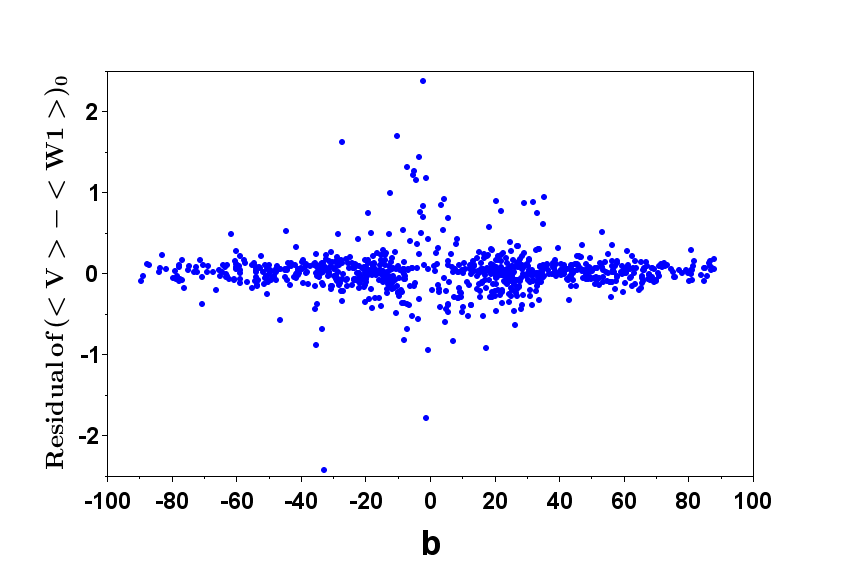}
    \caption{Plot of residual of $(V-W1)_0$ (equation~(\ref{eq:E16a})) as a function of $b$.}
    \label{fig:Residual}
\end{figure}

Fig.~\ref{fig:V_W1_Fe} shows the plot of equation~(\ref{eq:E16a}) for all 846 RRLs with linear fitting   with
parameters from~(\ref{eq:E16b}) superimposed, and Fig.~\ref{fig:Residual} shows the plot of its residuals as a function of Galactic latitude. 
The most interesting feature 
of Fig.~\ref{fig:Residual} is that residuals decrease sharply for $|b|\geq40^{0}$, which indicates that the \mbox{\cite{drimmel2003three}} 3D extinction model is 
accurate 
for high Galactic latitudes as suggested in \cite{dambis2013rr}. We therefore perform new fits based on  271 and 308 RR Lyraes located at $|b|>40^{0}$ (recall that in the case of the $(V-Ks)_0$
calibration we use only phase-corrected $K_s$-band data for the total  subsample): 
\begin{equation}
\begin{array}{rcl}
(V-Ks)_0&=&1.834(\pm0.028)+0.131(\pm0.018)[\\
         & & Fe/H]+2.33\log P_F \label{eq:E15c}\\
\end{array}
\end{equation}
with a scatter of $\sigma(V-Ks)_0=0.113$ (264 of 271 stars)
and
\begin{equation}
\begin{array}{rcl}
(V-W1)_0&=&1.865(\pm0.022)+0.108(\pm0.014)[\\
         & & Fe/H]+2.381\log P_F \label{eq:E16c}\\
\end{array}
\end{equation}
with a scatter of $\sigma(V-W1)_0=0.094$ (299 of 308 stars) and finally adopt these period-luminosity-colour relations for  RR Lyraes.

Hence 0.232~-~$b_K$~=~0.108 and 0.232~-~$b_W$~=~0.108, implying that $b_K$~=~0.101 and $b_{W}$~=~0.124. Given equations~(\ref{eq:E13}) and (\ref{eq:E14}) 
it follows from  this that
\begin{equation}
\langle M_{Ks}\rangle=a_K+0.101[Fe/H]-2.33\log P_F
\label{eq:PLZMKinit0}
\end{equation}
and
\begin{equation}
\langle M_{W1}\rangle=a_W+0.124[Fe/H]-2.381\log P_F,
\label{eq:PLZMWinit0}
\end{equation}
respectively.  To make the adopted PLZ relations maximally consistent with those inferred by \cite{dambis2013rr}
and  \cite{dambis2014mid}, respectively, we set the zero points $a_K$ and $a_W$ so as to obtain the same absolute magnitudes at the average metallicity and 
average logarithm of the fundamenalised period
for our RR Lyrae sample ($<[Fe/H]>$~=-1.39 and $<\log P_F>$~=~-0.277), so that:
\begin{equation}
\langle M_{Ks}\rangle=-0.809+0.101[Fe/H]-2.33\log P_F
\label{eq:PLZMKinit1}
\end{equation}
and
\begin{equation}
\langle M_{W1}\rangle=-0.789+0.124[Fe/H]-2.381\log P_F,
\label{eq:PLZMWinit1}
\end{equation}   
respectively and it is these relations that we use below to compute our initial RR~Lyrae distances for the statistical-parallax analysis. Note that 
the dependence of the $A_V$  reddening estimates for RR Lyraes at high Galactic latitudes, which are based on 
the 3D interstellar extinction model by 
\cite{drimmel2003three} on the adopted PLZ relation, is negligible as we found earlier by comparing the extinction values computed with RR Lyrae distances inferred using different
PLZ relations: the mean offset is of about +0.0002~$\pm$0.0001 with a scatter of 0.0016 \citep{dambis2013rr}.

 It is not safe to use extinction values derived from the 3D map at low Galactic latitudes, where the clumpy distribution of absorbing matter makes 
such estimates  rather inaccurate and too sensitive to distance errors. We therefore finally determined the interstellar extinction for 541 RRLs  that are located at low galactic 
latitudes ($|b|\leq 40^{0}$) from their $E(V-W1)$  (538 stars) or $E(V-Ks)$  (3 stars) colour excess based on period-metallicity-colour relations (\ref{eq:E15c}) and (\ref{eq:E16c}) above
by converting the latter  into  $E(B-V)$ and interstellar extinction values ($A_V, A_{Ks}, A_{W1}$) adopting the \cite{yuan2013empirical}   extinction
coefficients.

\begin{figure}
\includegraphics[width=\columnwidth]{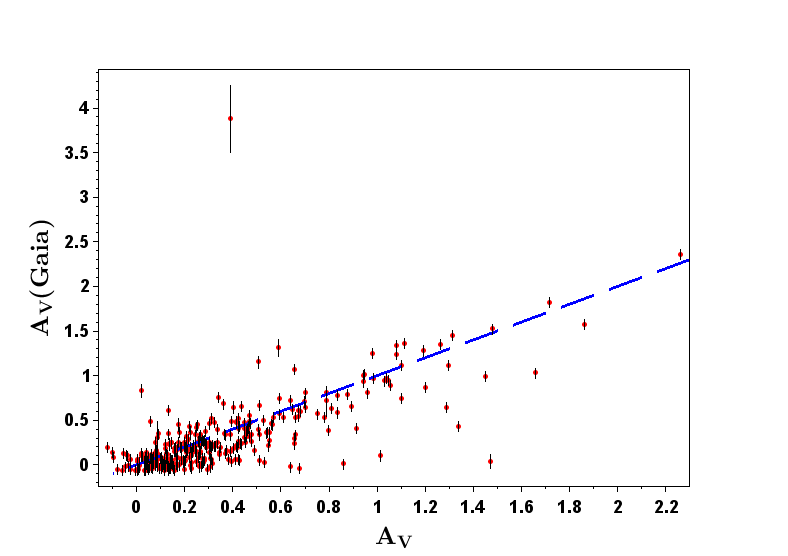}
    \caption{Comparison between the V-band interstellar extinction derived via our calibrated colour relations and \textit{Gaia} DR2.The broken line represents the bisector.}
    \label{fig:Av_vs_AG}
\end{figure}

In Fig.~\ref{fig:Av_vs_AG}, we compared our V-band interstellar extinction determined by using our calibrated colour relations for our sample RRLs with their available \textit{Gaia} DR2 values\citep{clementini2019gaia}. In this case, the values of $A_G$ in \textit{Gaia} G-band photometry were converted to $A_V$ using the extinction coefficient of \cite{chen2018three} ($R_G=2.5$) to be consistent with \cite{yuan2013empirical}  extinction coefficients. The lower left corner
of Fig.~\ref{fig:Av_vs_AG} shows the nice agreement existing between our interstellar extinction and \textit{Gaia} estimation for  lower values; on the
contrary, there is less agreement for larger extinction values which may be due to less information  obtained from the three broad \textit{Gaia} pass bands to estimate larger  extinction values \citep{anders2019photo}.  Hence, caution must be taken to use \textit{Gaia} DR2 extinction values as suggested by \cite{arenou2018gaia}. The final V-band interstellar extinction values ($A_V$) for our total sample RRLs are presented in column 9 of Table \ref{tab:table_1}.

\subsection{Final Database}
Using the data sources described above, we constructed a database for a total of 850 RRLs as presented in Table \ref{tab:table_1}  (the full version will be available in the CDS and  in the online version of the article--see Supporting Information). The
columns of this table are the following: (1) the GCVS name of the star;
(2) and (3) its J2000.0 equatorial coordinates $ \alpha $ and $ \delta $, respectively,
in decimal degrees; (4) the variability period in days; (5) the RRL type (AB or C); (6) and (7) the V-band  intensity-mean magnitude $<m_V>$ with its error and data source reference code  (Ref: 1 for our observation; 2 for \textit{Gaia} DR2 and 3 for \cite{dambis2013rr} catalog), respectively;  (8) the V-band
interstellar extinction $A_V $; (9) and (10) the metallicity [Fe/H] on the \cite{zinn1984globular} scale with its error and data source reference code  (Ref: 1 for our observation and 3 for \cite{dambis2013rr} catalog); (11) and (12) the Ks-band intensity-mean magnitude $<Ks>$ or \textit{2MASS} single epoch magnitude with its error $\sigma_{Ks}$  and data source reference code  (Ref: 7 for \textit{ 2MASS} single epoch observation, 8 for \cite{layden2019infrared}  and 9 for mean estimation in this paper and  \cite{dambis2013rr}), respectively; (13) and (14) the
W1-band intensity-mean magnitude  $<W1>$ or \textit{WISE} single epoch magnitude with its   errors $\sigma_{W1}$ and data source reference code (Ref: 3 for \cite{dambis2013rr} catalog, 4 for \cite{chen2018wide} catalog, 5 for \cite{gavrilchenko2014mid} catalog and 6 for  \textit{ALL WISE} single epoch magnitudes), respectively; (15) and (16) the average radial velocity $V_R$ with its
standard error $\sigma V_R$ (both in \,km\,s$^{-1}$) and data source reference code  (Ref: 1 for our observation and 3 for \cite{dambis2013rr} catalog),
respectively; (17)-(19)  the \textit{Gaia} DR2 proper motion in \,mas\,yr$^{-1}$ to the right ascension ($ \mu_\alpha $) and declination $(\mu_\delta)$ and parallax in mas with their corresponding standard errors, respectively.

\begin{table*}
	\caption{The first 10 lines of the total 850 Galactic field RRL variables database. This is a sample of the full version, which is
available in the CDS and the online version of the article (see Supporting Information).}
	\label{tab:table_1}
\begin{sideways}
\begin{tabular}{lllllllllllllllllll} 
\hline
 (1) &  (2)      &   (3)     &    (4)   &(5) &  (6)              & (7)& (8)    &   (9)          & (10) &             (11) & (12) &         (13)   &   (14)  \\
Name   & $\alpha $ & $\delta $ & Period & Type & $<V>$             & Ref& Av     & [Fe/H]         & Ref& $<Ks>$            & Ref& $<W1>$            & Ref\\
\hline                                                                                                                                                   
SW And & 5.92954   & 29.401022 & 0.4423 & RRAB &  9.712 $\pm$0.009 & 3  & 0.28408  & -0.38$\pm$  -  & 3  &  8.511 $\pm$0.009 & 8  &  8.464 $\pm$0.009 &  3 \\
XX And & 19.3642   & 38.950554 & 0.7228 & RRAB &  10.687$\pm$0.009 & 3  & -0.00359 & -2.01$\pm$  -  & 3  &  9.442 $\pm$0.027 & 8  &  9.378 $\pm$0.006 &  3 \\
ZZ And & 12.3952   & 27.02213  & 0.5545 & RRAB &  13.082$\pm$0.018 & 3  &  0.21469 & -1.58$\pm$  -  & 3  &  11.834$\pm$0.035 & 9  &  11.795$\pm$0.01  &  3 \\
XY And & 21.6768   & 34.06858  & 0.3987 & RRAB &  13.68 $\pm$0.019 & 3  &  0.26325 & -0.92$\pm$  -  & 3  &  12.892$\pm$0.04  & 9  &  12.617$\pm$0.005 &  3 \\
BK And & 353.775   & 41.102886 & 0.4216 & RRAB &  12.97 $\pm$0.037 & 3  &    0.397 & -0.08$\pm$  -  & 3  &  11.663$\pm$0.04  & 9  &  11.614$\pm$0.008 &  3 \\
BK Ant & 146.062   & -39.6613  & 0.5166 & RRAB &  11.799$\pm$  -   & 2  &  0.13406 & -0.519$\pm$0.16 & 1  &  10.614 $\pm$0.04 & 9  &  10.547$\pm$0.023 &  4 \\
BN Ant & 149.275   & -39.2906  & 0.5359 & RRAB &  12.488$\pm$  -   & 2  & 0.31463  & -1.25$\pm$0.16 & 1  &  11.146$\pm$0.04 & 9  &  11.107$\pm$0.023 &  4 \\
RT Ant & 143.064   & -25.1997  & 0.553  & RRAB &  14.198$\pm$  -   & 1  &  0.29522 & -1.48$\pm$0.16 & 1  &  12.878$\pm$0.038 & 7  &  12.828$\pm$0.024 &  6 \\
SS Ant & 144.795   & -26.5519  & 0.4837 & RRAB &  13.526$\pm$  -   & 2  &  0.28575 & -1.42$\pm$0.16 & 1  &  12.582$\pm$0.04 & 9  &  12.463$\pm$0.024 &  5 \\
UZ Ant & 166.334   & -38.9462  & 0.6003 & RRAB &  14.0  $\pm$  -   & 2  &  0.24006 & -1.75$\pm$0.16 & 1  &  12.361$\pm$0.042 & 7  &  12.354$\pm$0.023 &  5 \\
\hline
(1)  &  (15)          &(16) &  (17)          &       (18)    &  (19)         \\
Name   & $V_R$          & Ref&$\mu_\alpha\pm\sigma \mu_\alpha$ & $\mu_\delta\pm\sigma \mu_\delta$ & $\varpi\pm\sigma\varpi$\\
SW And & -21.0$\pm$1.0  &  3 &  -6.8 $\pm$0.1  & -19.3$\pm$0.3 &  1.7797$\pm$0.1636 \\
XX And & 0.0  $\pm$1.0  &  3 &  58.1 $\pm$0.1  & -32.9$\pm$0.1 &  0.695 $\pm$0.0463 \\
ZZ And & -13.0$\pm$53.0 &  3 &  31.6 $\pm$0.1  & -17.5$\pm$0.0 &  0.3148$\pm$0.0273 \\
XY And & -64.0$\pm$53.0 &  3 &  11.3 $\pm$0.1  & -9.0 $\pm$0.1 &  0.338 $\pm$0.0429 \\
BK And & -17.0$\pm$7.0  &  3 &  6.3  $\pm$0.0  & -1.4 $\pm$0.0 &  0.445 $\pm$0.0281 \\
BK Ant &  30.4$\pm$10.5  &  1 &  0.3  $\pm$0.0  & -3.3 $\pm$0.0 &  0.6422$\pm$0.0304 \\
BN Ant & 258.4$\pm$22.8  &  1 &  24.1 $\pm$0.0  & -26.8$\pm$0.0 &  0.4264$\pm$0.0278 \\
RT Ant & 396.0$\pm$24.0  &  1 &  -4.8 $\pm$0.1  & -2.3 $\pm$0.1 &  0.2202$\pm$0.0327 \\
SS Ant & 45.7  $\pm$18.3  &  1 &  9.8  $\pm$0.1  & -15.4$\pm$0.1 &  0.1652$\pm$0.033  \\
UZ Ant & 515.7$\pm$11.2  &  1 &  -17.6$\pm$0.0  & -5.8 $\pm$0.0 &  0.1884$\pm$0.0244 \\
\hline
	\end{tabular}
\end{sideways}
\end{table*}
\section{Method}
\label{sec:Method} 
As we pointed out earlier, this research uses the bimodal version of maximum-likelihood statistical parallax method to calibrate the zero points of LZ and PLZ relations of RRLs  in visual and IR bands, and to estimate the kinematical parameters of RRLs population in the disk and halo of our Galaxy.  Here, we present the basic ideas and notation of the method as described in section 2.4 of \cite{dambis2009}; for detailed step-by-step instructions see the papers by \cite{hawley1986absolute} and \cite{dambis2009}.

We begin by computing the assumed distances in kpc for our sample RRLs using 
\begin{equation}
\label{eq:E2m}
D_{A}=10^{0.2(\langle m_{x}\rangle-A_{x} - M_{A,x})-2}
\end{equation}
where $\langle m_{x}\rangle$, $M_{A,x}$ and $A_{x}$ are the apparent mean magnitude, the adopted absolute magnitude and interstellar extinction in x band, respectively. For reasons that are mentioned in section \ref{sec:introd} and additional explanations in the next section \ref{sec:Results},  we estimated $M_A$ for almost all of our sample RRLs by adopting the above calibrated W1-band PLZ relation of \cite{dambis2014mid} (eq.~\ref{eq:PLZMWinit1}).  However, for  four RRLs that have no  measured $\langle m_{W1}\rangle$, we adopted Ks- band PLZ relation (eq.~\ref{eq:PLZMKinit1}) which is consistent with the above W1-band PLZ relation. Thus, our aim is to use the statistical-parallax
technique to determine the absolute-magnitude correction for the adopted PLZ  relations  ($\Delta M; M_{true}=M_{A}+\Delta M$), which is related to the inverse distance-scale correction factor ($f = D_{A} / D$; the assumed distance divided by the real or true distance) in such a way that $ f=10^{0.2 \times\Delta M}$.

Using the assumed distance-$D_{A}$ a residual velocity ($\bmath{\nu}$) for a given local homogeneous sample, which is equal to the observed velocity minus the velocity implied by the adopted model, can be calculated:
\begin{equation}
 \bmath{\nu} =|\bmath{D_{A}}|\bmath{\mu_{obs}}+V_{R(obs)}\hat{\bmath{r}}-[f\bmath{\Pi} +\bmath{\hat{r}\hat{r}}'](\bmath{\bar{V}-\bar{V}_\odot}),
\end{equation}
where $\bmath{\mu_{obs}}$,  $\bmath{V_{R(obs)}}$  $\bmath{\bar{V}}$ and $\bmath{\bar{V}_\odot}$ are observed proper motion, radial velocity, relative bulk (average) velocity of the sample star and the Sun with respect to the adopted reference frame, respectively; $\bmath{\hat{r}}$- is the unit vector pointing radially towards the star and $\bmath{\hat{r}}'$- is its transpose; $\bmath{\hat{r}\hat{r}}'$ is the matrix of projection on to the line of sight; $\bmath{\Pi=U-\hat{r}\hat{r}}'$ is the matrix of projection
on to the sky plane, and \textbf{U} is the unit matrix. Here, $ \bmath{\bar{V}-\bar{V}_\odot} $ is one of the unknown model parameters that we need to determine.

Thus, the matrix of covariance (\textbf{M}), which is the covariances tensor of $\bmath{\nu} $, can be computed  in terms of the assumed distance of the star, the distance-scale correction factor f, the random error of the distance-scale parameter ($\delta f$), the observational errors of proper motions ($\delta\mu$) and the standard error of the radial velocity $(\delta V_R)$:
\begin{equation}
\begin{array}{rcl}
 \textbf{M} &=& \displaystyle cov(\bmath{\nu})=\langle \delta\bmath{\nu}\delta\bmath{\nu}' \rangle=|\bmath{D_{A}}|^{2}\langle \delta\bmath{\mu}\delta\bmath{\mu}'\rangle+  \\
 & &\displaystyle(\hat{\bmath{r}}\hat{\bmath{r}}')\delta V_{R}^{2} +[f\bmath{\Pi} +\bmath{\hat{r}\hat{r}}']\bmath{\Sigma}[f\bmath{\Pi}+\bmath{\hat{r}\hat{r}}']'+ \\
 & &\displaystyle\delta f^{2} \bmath{\Pi}[(\bmath{\bar{V}-\bar{V}_\odot)(\bar{V}-\bar{V}_\odot)}'+\bmath{\Sigma}]\bmath{\Pi}'
\end{array},
\end{equation}
where $\bmath{\Sigma}$ is the covariance tensor of true residual velocities at the location of the star (the velocity ellipsoid), and this is also another unknown model parameter that we are supposed to determine. 

If we assume that residual velocities $\bmath{\nu} $ obey a three-dimensional Gaussian distribution with
zero mean and a covariance tensor \textbf{M}  in the local homogeneous sample, then the likelihood of obtaining all observations simultaneously for N-stars can be expressed as
\begin{equation}
\begin{array}{rcl}
 L &=& \displaystyle\prod_{i=1}^{N}prob(\bmath{\nu}_i)\\
   &=& \displaystyle\prod_{i=1}^{N} \lbrace(  2\pi)^{-\frac{3}{2}} |\textbf{M}|_{i}^{-\frac{1}{2}}\exp(-0.5\bmath{\nu}'_i\textbf{M}_{i}^{-1}\bmath{\nu}_i)\rbrace,
\end{array}
\end{equation}
where $|\textbf{M}|_{i}$ and $\bmath{\nu}'_{i}$ denotes the determinant of  the covariance tensor and the transpose of residual velocity vector for specific star i, respectively. Hence, the distribution function
has the meaning of the probability density for the residual
velocity of a particular star, and given that stars are distributed
independently of each other in the velocity space, their N-
point distribution function is equal to the product of functions for all stars of the sample. Therefore, the logarithm of the likelihood of obtaining all observations simultaneously becomes
\begin{equation}
\begin{array}{rcl}
 \ln L &=& \displaystyle\sum_{i=1}^{N} \ln prob(\bmath{\nu}_i) \\
     &=& \displaystyle-0.5\sum_{i=1}^{N} (\ln |\textbf{M}|_{i}+\bmath{\nu}_{i}^{'}\textbf{M}^{-1}_{i}\bmath{\nu_{i}})+\frac{3}{2}(2\pi)N
\end{array} 
\end{equation}

 However, our sample is not homogeneous as it contains  two purportedly homogeneous
subsamples (Disk and Halo), each of which obeys its own three-dimensional residual velocity  distribution and each object may  priorly belong to any of the
two subsamples. In this case, the residual velocity distributions for any specific star i becomes
\begin{equation}
\begin{array}{rcl}
  prob(\bmath{\nu}_i) &=& \displaystyle \alpha\times (2\pi)^{-\frac{3}{2}}|\textbf{M}_{1}|^{-\frac{1}{2}}\exp(-0.5\bmath{\nu}'_{1}\textbf{M}_{1}^{-1}\\
  & &\bmath{\nu}_{1})+(1-\alpha)\times(2\pi)^{-\frac{3}{2}}|\textbf{M}_{2}|^{-\frac{1}{2}}\\
  & &\exp(-0.5\bmath{\nu}'_{2}\textbf{M}^{-1}_{2}\bmath{\nu}_{2}), \label{eqn:E21} 
\end{array}
\end{equation}
where $\alpha$ and $1-\alpha$ are the  number fractions of the halo (j=1) and disk (j=2) subsamples (prior probability for a single star i to be in halo and disk subsamples) and these are unknown parameters that we need to  estimate; $\bmath{\nu}_{j}$ are the residual velocities of the specified star i with respect to $j^{th}$ subsample, which can also be expressed as
\begin{equation}
 \bmath{\nu}_j =|\bmath{D_{A}}|\bmath{\mu}_{obs}+V_{R(obs)} \,\,\bmath{\hat{r}}-[f\bmath{\Pi} +\bmath{\hat{r}\hat{r}}'](\bmath{\bar{V}}_j-\bmath{\bar{V}}_{\odot j});
\end{equation}
 here, $\bmath{\bar{V}}_j$ and $\bmath{\bar{V}}_{\odot j}$ are the average velocity of the $j^{th}$ subsample stars and the Sun at the specified star i location with respect to the  adopted reference frame, respectively, and $\textbf{M}_{j},j=1,2$ are the corresponding  covariance tensors. 

When we express the likelihood function ($\ln L$) of the bimodal RRL population distribution function (equation \ref{eqn:E21}) in terms of the observable quantities and the model parameters to be determined, the likelihood function 
depends on 14 unknown model parameters: the three components of the relative bulk motion of the halo, ($U_{0}$, $V_{0}$, $W_{0}$)$_{Halo}$ and disk, ($U_{0}$, $V_{0}$, $W_{0}$)$_{Disk}$, RRL subsamples with respect to the Sun (the reflex solar motion with respect to the sample stars); the three diagonal components of the velocity ellipsoid of the halo, $(\sigma V_{R}$, $\sigma V_{\phi}$, $\sigma V_{\theta})_{Halo}$ and disk, $(\sigma V_{R}$, $\sigma V_{\phi}$, $\sigma V_{\theta})_{Disk}$, RRL components; the inverse
distance-scale correction factor (f), and the fraction of RRLs that belong to halo component ($\alpha$).  To this end, we adopt the peculiar velocity of sun to be (+9, +12, +7) km/s   \citep{mihalas1981galactic}  and the  local
standard of rest velocity to be 220 km/s  \citep{kerr1986review}  in the Cartesian Galactic system.

 Thus,  the best values of these 14 model parameters were determined according to the principle of maximum likelihood by maximizing $\ln L$, which is equivalent to minimizing $-\ln L$, using the numerical method of minimization without evaluation of derivatives as developed by \cite{pshenichnyi1976numerical}. 
 We calculate the standard errors for these inferred model parameters by  the following formula:
\begin{equation}
\sigma^2_i=\frac{\int(par_i-<par_i>)^2L(par_1,...,par_N) \! \mathrm{d}par_1...\mathrm{d}par_N}{\int(L(par_1,...,par_N) \! \mathrm{d}par_1...\mathrm{d}par_N}
\end{equation} 

To check and (if necessary) correct the above results for any   possible biases, like not considering the effect of sample selection and uncertainty in the adopted distance, we followed a similar
formalism as \cite{dambis2013rr}.  Thus, we generated an entire ensemble of  2500 simulated samples consisting of stars with exactly the same sky coordinates as in our data Table \ref{tab:table_1} by imposing the above solution and adding random normally distributed noise to all simulated observables with
the variances in accordance with the corresponding standard errors of actual input data. We then applied to each of the simulated samples
the same maximum-likelihood algorithm as we used for the corresponding actual sample to determine the bias corrections for all inferred parameters. As
we show in the next next section (\ref{sec:Results}), where we present the results, the corrections prove to be very small compared to the standard errors of
the inferred parameters. 

\section{Results}
\label{sec:Results} 
\subsection{Kinematic parameters for the disc and halo population of the Milky Way RRLs}
Our results for 14 statical parallax parameters using the maximum-likelihood version of the statistical parallax 
method, as described 
in section~\ref{sec:Method}, applied on the entire 850 RRL sample without 
priorly assigning any particular star to the thick disc or halo population, are listed in 
Table~\ref{tab:table_2}. 
Its columns give the results for the following parameters: the fraction of stars of our sample belonging to the corresponding population; the 
components of the mean mean heliocentric velocity $(U_0 ,V_0 , W_0 )$ and velocity dispersion $(\sigma V_R, \sigma V_\phi , \sigma V_\theta)$ 
along the corresponding axes directed toward the
Galactic center, in the sense of Galactic rotation, and toward the North 
Galactic Pole, 
respectively. The first two rows give the initially inferred parameter values for the halo population and their standard errors, respectively;
The next two rows (3 and 4) give the parameter bias values $\delta_{par}$ and their standard errors, respectively, estimated from 2500 randomly simulated samples
with the inferred kinematics and simulated observational errors superimposed;
the next two rows (5 and 6), which are printed in bold, give the bias-corrected parameter values and their standard errors, respectively. The next
two rows (7 and 8) give the scatter of the parameter values estimated for the set of  2500 simulated samples and the ratios of the scatter to
the estimated standard deviation (these ratios allow us to assess how well does our solution reproduce the parameter errors), respectively. 
These eight rows are followed by a blank line and another eight rows with the results for the disc population (their layout is the same as for
the halo population). The last five rows give the initial absolute-magnitude correction $\Delta M_{}$ with its estimated standard error,
the $\Delta M_{}$ bias correction $\delta \Delta M_{}$ with its standard error as estimated from 2500 simulated samples, 
the bias-corrected $\Delta M_{}$ with its standard error (printed in bold), the scatter of $\Delta M_{}$ estimates for 2500 simulated samples, and the ratio of this 
scatter to the estimated standard error.

\begin{table*}
	\centering
	\caption{Kinematical parameters and W1-band absolute-magnitude corrections of Galactic field RRL using bimodal statistical 
parallax solution on the entire 850 RRL sample with bias correction based on 2500 simulated samples}
	\label{tab:table_2}
	\begin{tabular}{lrrrrrrr} 
		\hline
		Population & Fraction of the sample & $U_0  $ & $ V_0 $ & $W_0$ & $\sigma V_R$ & $\sigma V_\phi$ &  $\sigma V_\theta$\\
		  & $\left(  \alpha\right)$   \& $\left( 1-\alpha \right)$  &  &  & $\left( kms^{-1}\right) $&  & & \\
		\hline  
		 Halo     & 0.805       & -10.08        & -220.54         & -10.49          & 141.62          & 107.27          & 107.08     \\
		          & $\pm 0.017$ & $\pm 5.28$   & $\pm 6.21$      & $\pm 4.22$      & $\pm 4.70$      & $\pm 3.70$      & $\pm 3.75$ \\
  Bias $\delta_{par}$     & -0.001       & 0.00        & -0.33           & +0.01           & -0.28           & -0.60           & -0.18      \\
                          & $\pm 0.001$ & $\pm 0.10$   & $\pm 0.13$      & $\pm 0.09$      & $\pm 0.09$      & $\pm 0.08$      & $\pm 0.07$ \\
{\bf Bias corrected   }   &{\bf 0.804 } & {\bf -10.08}  & {\bf -220.87}   & {\bf -10.48 }   & {\bf 141.34}    & {\bf 106.67}    & {\bf 106.90 }    \\
{\bf parameters}    & \mbox{\bf $\pm$ 0.017} & \mbox{\bf $\pm$ 5.28}  & \mbox{\bf $\pm$ 6.21}& \mbox{\bf $\pm$ 4.22}& \mbox{\bf $\pm$ 4.70}& \mbox{\bf $\pm$ 3.70}& \mbox{\bf $\pm$ 3.75} \\
       Scatter(par)    & 0.017       & 5.17         & 6.36            & 4.29            & 4.42            & 3.80            & 3.67       \\
Scatter(par)/$\sigma_{par}$ & 1.00   & 0.98         & 1.02            & 1.02            & 0.94            & 1.03            & 0.98       \\
\\
		 Disc     & 0.195       & -21.91       & -46.29          & -15.15          & 54.03           & 36.54           & 27.96\\
		          & $\pm 0.017$ & $\pm 5.14$   & $\pm 4.12$      & $\pm 2.76$      & $\pm 4.40$      & $\pm 3.53$      & $\pm 2.56$\\
  Bias $\delta_{par}$     & +0.001       & -0.16        & -0.06           & +0.09           & -0.10           & -0.17           & -0.23      \\
                          & $\pm 0.001$ & $\pm 0.10$   & $\pm 0.08$      & $\pm 0.06$      & $\pm 0.09$      & $\pm 0.06$      & $\pm 0.05$ \\
{\bf Bias corrected    }  & {\bf 0.196 }& {\bf -22.07} & {\bf -46.35 }   & {\bf -15.06  }  & {\bf 53.93    } & {\bf 36.37    } & {\bf 27.73     }\\
{\bf parameters}    & \mbox{\bf $\pm$ 0.017} & \mbox{\bf $\pm$ 5.14}  & \mbox{\bf $\pm$ 4.12}& \mbox{\bf $\pm$ 2.76}& \mbox{\bf $\pm$ 4.40}& \mbox{\bf $\pm$ 3.53}& \mbox{\bf $\pm$ 2.56} \\
       Scatter(par)    & 0.017       & 4.93         & 3.96            & 2.88            & 4.54            & 3.25            & 2.55       \\
Scatter(par)/$\sigma_{par}$ & 1.00   & 0.96         & 0.96            & 1.04            & 1.03            & 0.92            & 1.00       \\
\\
       $Delta M_{}$                          & +0.011$\pm$ 0.056        \\  
$\delta\Delta M_{}$                          & -0.002$\pm$ 0.001        \\  
{\bf Bias corrected $\Delta M_{}$}                      & {\bf 0.009$\pm$ 0.056}        \\  
$scatter_{\Delta M_{}}$                      & 0.057 &\\
$scatter_{\Delta M_{}}$/$\sigma \Delta M_{}$ & 1.02\\
		\hline
	\end{tabular}
\end{table*}

\begin{figure}
\includegraphics[width=\columnwidth]{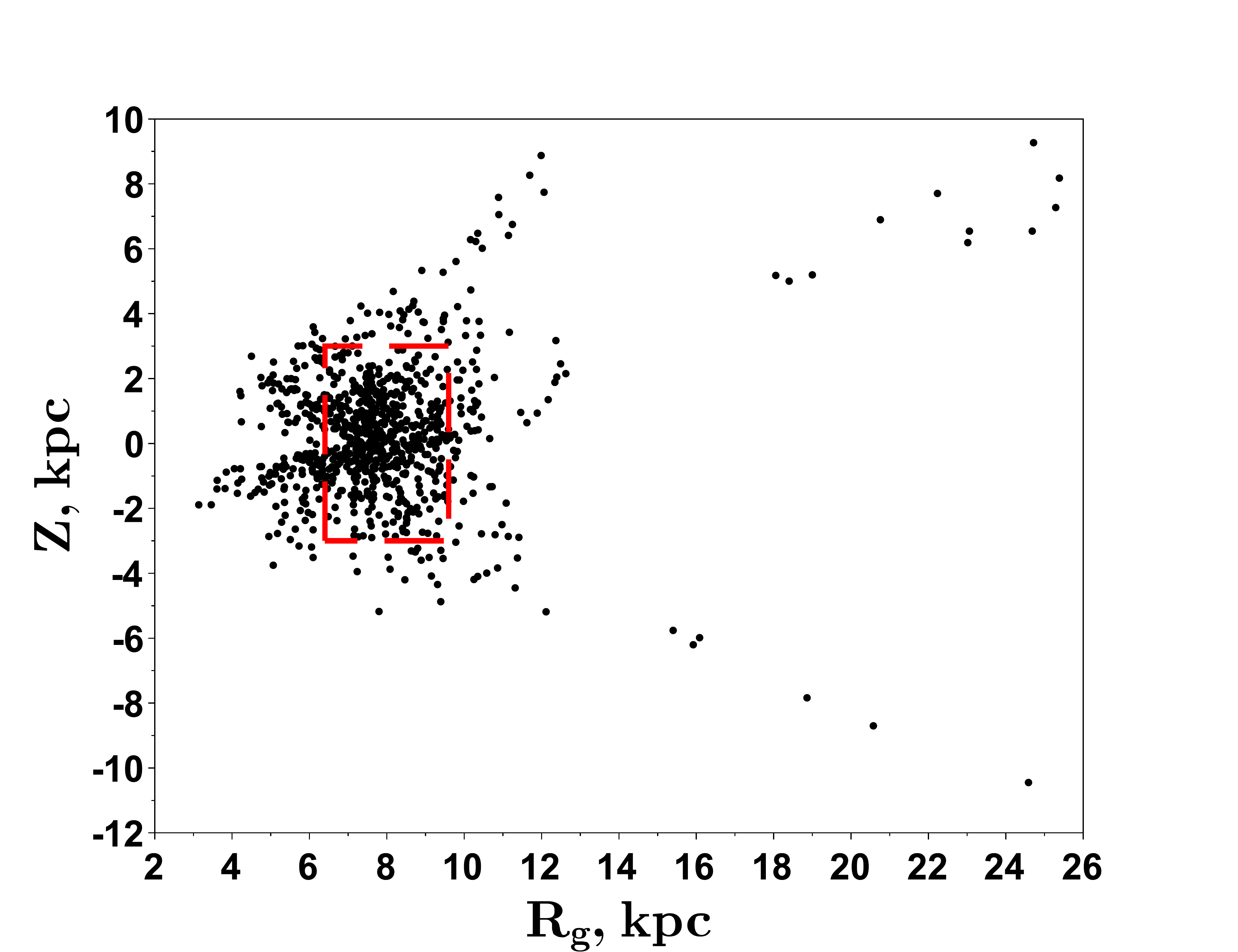}
    \caption{ Distribution of RR Lyraes of our sample in  the $R_g$-$Z$ coordinates. The red rectangle shows the domain of RR Lyraes selected for the $R_g$ and $Z$-
clipped solution in Table~\ref{tab:table_3}}
    \label{fig:distance_z}
\end{figure}

However, our result in Table~\ref{tab:table_2}  relies on one assumption 
that is not quite correct: the velocity distributions  are assumed to be position independent. This argument is not true  over the  broad interval of the Galactocentric distances and heights above the Galactic plane spanned by our RR Lyrae type stars 
shown in Figure~\ref{fig:distance_z}. To 
 eliminate or at least reduce   this effect, we recomputed the result by applying our procedure to the subsample of  533 RRLs with close to solar Galactocentric distances, $ 6.4.0\leq R_g \leq 9.6 $ kpc ( 63 \% of our total 850 RRLs sample) and present the result  in Table~\ref{tab:table_3} (its layout is identical to that of Table~\ref{tab:table_2}).

\begin{table*}
	\centering
	\caption{Kinematical parameters and W1-band absolute-magnitude corrections of Galactic field RRLs from bimodal statistical parallax solution based on
         533 field RRLs with close to solar  Galactocentric distances ($6.4 \leq R_{g} \leq 9.6$ kpc) located close to the Galactic plane ($|z| \leq 3.0$ kpc)) with bias correction based on 2500 simulated samples}
	\label{tab:table_3}
	\begin{tabular}{lrrrrrrr} 
		\hline
		Population & Fraction of the sample & $U_0  $ & $ V_0 $ & $W_0$ & $\sigma V_R$ & $\sigma V_\phi$ &  $\sigma V_\theta$\\
		  & $\left(  \alpha\right)$   \& $\left( 1-\alpha \right)$  &  &  & $\left( kms^{-1}\right) $&  & & \\
		\hline  
		 Halo     & 0.763       & -15.68       & -218.62         & -6.38           & 152.90          & 106.31          & 101.35     \\
		          & $\pm 0.023$ & $\pm 7.14$   & $\pm 7.37$      & $\pm 5.06$      & $\pm 6.59$      & $\pm 4.30$      & $\pm 4.42$ \\
Bias   $\delta_{par}$     & -0.001      & -0.24        & -0.37           & -0.05           & -0.14           & -0.61           & -0.19      \\
                          & $\pm 0.001$ & $\pm 0.15$   & $\pm 0.15$      & $\pm 0.10$      & $\pm 0.13$      & $\pm 0.09$      & $\pm 0.08$ \\
{\bf Bias corrected   }   &{\bf 0.762 } & {\bf -15.92} & {\bf -218.99}   & {\bf -6.43 }    & {\bf 152.76}    & {\bf 105.70}    & {\bf 101.16 }    \\
{\bf parameters}    & \mbox{\bf $\pm$ 0.022} & \mbox{\bf $\pm$ 7.37}  & \mbox{\bf $\pm$ 5.06}& \mbox{\bf $\pm$ 6.59}& \mbox{\bf $\pm$ 4.30}& \mbox{\bf $\pm$ 4.42}& \mbox{\bf $\pm$ 4.61} \\
       Scatter(par)    & 0.022       & 7.49         & 7.49            & 5.11            & 6.28            & 4.71            & 4.19       \\
Scatter(par)/$\sigma_{par}$ & 0.95   & 1.05         & 1.02            & 1.01            & 0.95            & 1.09            & 0.95       \\
\\
		 Disc     & 0.237       & -19.05       & -45.66          & -14.01          & 49.07           & 38.62           & 25.45\\
		          & $\pm 0.023$ & $\pm 5.19$   & $\pm 4.92$      & $\pm 2.69$      & $\pm 4.46$      & $\pm 3.93$      & $\pm 2.54$\\
Bias   $\delta_{par}$     & +0.001       & -0.15        & -0.10           & +0.01           & -0.30           & -0.25           & -0.18      \\
                          & $\pm 0.001$ & $\pm 0.10$   & $\pm 0.09$      & $\pm 0.06$      & $\pm 0.09$      & $\pm 0.07$      & $\pm 0.05$ \\
{\bf Bias corrected    }  & {\bf 0.236 }& {\bf -19.20} & {\bf -45.76 }   & {\bf -14.00  }  & {\bf 48.77    } & {\bf 38.37    } & {\bf 25.27     }\\
{\bf parameters}    & \mbox{\bf $\pm$ 0.023} & \mbox{\bf $\pm$ 5.19}  & \mbox{\bf $\pm$ 4.92}& \mbox{\bf $\pm$ 2.69}& \mbox{\bf $\pm$ 4.46}& \mbox{\bf $\pm$ 3.93}& \mbox{\bf $\pm$ 2.54} \\
       Scatter(par)    & 0.022       & 5.00         & 4.40            & 2.85            & 4.51            & 3.59            & 2.53       \\
Scatter(par)/$\sigma_{par}$ & 0.95   & 0.96         & 0.89            & 1.06            & 1.01            & 0.91            & 1.00       \\
\\
       $Delta M_{}$                          & -0.011$\pm$ 0.068        \\  
$\delta\Delta M_{}$                          &-0.004$\pm$ 0.001        \\  
{\bf Bias corrected $\Delta M_{}$}                      & {\bf -0.015$\pm$ 0.068}        \\  
$scatter_{\Delta M_{}}$                      & 0.068 &\\
$scatter_{\Delta M_{}}$/$\sigma \Delta M_{}$ & 1.00\\
		\hline
	\end{tabular}
\end{table*}

 As it is evident from Tables~\ref{tab:table_2} and \ref{tab:table_3}, the adopted maximum-likelihood statistical-parallax method yields
practically bias-free results (the parameter biases never exceed 9\% of the corresponding standard errors). Moreover, the error estimates
also prove to be quite reliable as shown by the ratio of the scatter of the parameter values inferred for mock samples to the standard
errors returned by the statistical-parallax method: these ratios are always contained in the interval from  0.89 to 1.06.

 We thus adopt parameters given by the solution applied on the solar neighborhood sample (Table~\ref{tab:table_3}) as our final result. Examination of Table~\ref{tab:table_3} shows that  disk RRLs at close-to-solar Galacticentric 
distances rotate rapidly with small dispersion, ($ V_0=187.2,  \sigma V_\phi=38.4)$ km/s, while 
 halo RRLs  with close to solar $R_g$ rotate
slowly with a large dispersion, ($ V_0=14.0, \sigma V_\phi=105.7$) km/s,   with respect to stationary Galactic center. 
These  results  agree with the estimates of \cite{carollo2010structure} for the thick Disk ($ V_0=182, \sigma V_\phi=51$)km/s and  inner halo stars ($V_0=7, \sigma V_\phi=95$)km/s (see their table 5). Furthermore,  our results are also consistent with a kinematically
hot ($V/\sigma<< $1) halo and a kinematically cold  ($V/\sigma \sim 1$) disk. 

We also see that our result for halo RR Lyrae is  strongly radially anisotropic ($ \beta= 0.54 $) 
as suggested by \cite{wegg2019gravitational} (see their fig 7, 8,9) form their study of  the  
RRL kinematics as a tracer of inner Galactic halo using \textit{Gaia} DR2. Our result of $ \beta= 0.54 $ 
 for solar-distance RRLs is less than the $ \beta=0.7-0.8 $ measured by \cite{smith2009kinematics,bond2010milky,belokurov2018co}, but \cite{belokurov2018co} show that the local
anisotropy of the stellar halo depends strongly on metallicity.

For comparison, we present in Table~\ref{tab:table_4} the kinematic results of some previous statistical parallax studies for halo RRLs. As  it is apparent from this table, our kinematic results for halo RRLs are in good agreement with previous works  while providing a  remarkable improvement in terms of errors  and are more accurate and precise than any other study of this kind so far. The improvement 
in precision  is due to the large number of RRLs used (as shown in the second column of Table~\ref{tab:table_4}) with 
accurate radial velocities and accurate  \textit{Gaia}  DR2 proper motions, which translate 
into accurate tangential velocities when multiplied by the  calibrated distance.
\begin{table*}
	\centering
	\caption{Kinematic results of some previous statistical parallax studies for their halo RRLs}
	\label{tab:table_4}
	\begin{tabular}{lccccccr} 
		\hline
		Reference & $N_{stars} $   & $U_0  $ & $ V_0 $ & $W_0$ & 
$\sigma V_R$ & $\sigma V_\phi$ &  $\sigma V_\theta$  \\
		\hline
\cite{hawley1986absolute} & 77  & 21$\pm$19 & -184$\pm$17	& -4$\pm$11 & 
166$\pm$76 & 114$ \pm$52 &91$\pm$40 \\
\cite{layden1996absolute} & 162  & -9$\pm$14 & -210$\pm$12 & -12$\pm$8 & 
168$\pm$13 & 102$\pm$8 & 97$\pm$7 \\
 \cite{tsujimoto1997absolute} & 99 & -12$\pm $17 & -200$\pm$11 & 1$\pm$3 & 
160$\pm$13 & 104$\pm$9 & 86$\pm$7 \\
 \cite{dambis2009} & 364 & -12$\pm$10 & -217$\pm$9 & -6$\pm$6 & 167$\pm$9 &
86$\pm$6 & 78$\pm$5  \\
 \cite{dambis2013rr} & 336 & -7$\pm$9 & -214$\pm$10 & -10$\pm$6 & 153$\pm$9 & 
101$\pm$6 & 96$\pm$5 \\
 This work &  533  & -16$\pm$7 & -219$\pm$7 &  -6$\pm$5 & 153$\pm$7 & 106$\pm$4
& 101$\pm$4 \\
		\hline
	\end{tabular}
\end{table*}

\subsection{ Zero-point calibration of LZ and PLZ relations based on statistical parallax  analysis }
 Our statistical parallax analysis of RRLs at close-to-solar  Galactocentric distances and located within 3~kpc from the Galactic plane 
as described above  yields an absolute magnitude correction of  $\Delta M_{}=-$0.015$\pm$0.068 (the last bold row of Table~\ref{tab:table_3}), 
 and our final calibrated PLZ relation in the $W1$ band and on \cite{zinn1984globular} metallicity scale becomes 

\begin{equation}
\begin{array}{rcl}
\langle M_{W1}\rangle &=& -0.824(\pm0.068)+0.124(\pm0.023)[Fe/ \\
&  &H]_{}-2.381(\pm0.097) \log P_F \label{eq:1}
\end{array}
\end{equation}
The implied zero point correction for PLZ relation in Ks-band and for LZ relation in visual V-band  results
\begin{equation}
\begin{array}{rcl}
\langle M_{Ks} \rangle &=& -0.804(\pm0.083)+0.101(\pm0.027)[Fe/\\
 & &H]_{}-2.33\log P_F \label{eq:5}\\
\end{array}
\end{equation}
and
\begin{equation}
\begin{array}{rcl}
\langle M_V\rangle &=& 1.041(\pm0.077)+0.232(\pm0.020)[Fe/H]_{}
\label{eq:3}\\
\end{array}
\end{equation}
, respectively.

\subsubsection{ Comparison of PLZ and LZ relations with earlier works } 
In resent years, many different authors have
calibrated the MIR  WISE $W1$-band PLZ relation for RRLs  both empirically \citep[e.g.][]{dambis2013rr,dambis2014mid,
sesar2017probabilistic,clementini2017gaia,muraveva2018rr}
and theoretically \citep[e.g][]{neeley2017new}. A brief summary of most of the previous studies are shown in table 3 of \cite{muraveva2018carnegie}. According to these studies, the value of the period slope varies from $-2.247\pm 0.018$ \citep{neeley2017new} to -2.470$\pm$0.074 \citep{sesar2017probabilistic};  the metallicity slope ranges from $0.088\pm0.026$ \citep{dambis2013rr} to $0.18\pm0.03$ \citep{neeley2017new}, and the zero point varies from $-0.79\pm0.07 $ \citep{neeley2017new} to $ -0.91\pm0.36$  \citep{muraveva2018rr}.

All empirically estimated coefficients and zero points for  the $W1$-band PLZ relation  differ in the adopted  metallicity scale and the zero point calibration methods, and  this makes 
comparison of our results with those of other researchers difficult. 
Regardless of this,  both our adopted coefficients and corrected zero point for W1-band PLZ relation marginally agree with all previous  empirical  and theoretical studies. However, our estimated metallicity  is  stepper than our previous measurement \citep{dambis2014mid} which had been made on single metallicity populations (i.e. in globular clusters). Moreover,  our calibrated zero point is consistent with previous statistical parallax studies, and it is slightly  brighter ($ \sim0.07$ mag) than the estimate obtained by \cite{muraveva2018rr} using other method.

 The zero-point of our  $M_V$-[Fe/H]  relation  at [Fe/H]=-1.5 ($M_V=+0.69 \pm0.08$) is consistent with all previous statistical
parallax results \citep[see][and
references therein]{dambis2013rr}  except \cite{kollmeier2013}  ($M_{V,RRC}=+0.52\pm0.11$ at $[Fe/H]=-1.59$ ), who calibrated LZ for RRC type variables.  It is also consistent with the value of $M_V =0.65\pm0.1$ mag obtained for the same metallicity by \cite{muraveva2018rr}, who applied Bayesian approaches  to 23 stars of our sample with homogeneous metallicity and \textit{Gaia} DR2 parallaxes to 
 model the LZ relation adopting the -0.057 mas parallax zero -point offset.

Like the visual $M_V-[Fe/H]$ relation, there  is extensive literature on the NIR Ks- band PLZ relation for RRLs addressing it either  from empirical \citep[e.g see][and
references therein]{clementini2017gaia,neeley2019standard,braga2018rr} 
or theoretical \citep[e.g.][]{bono2003pulsational,catelan2004rr,marconi2015new} points of view. A comprehensive list of most of the 
 published studies without the  present ones are presented in table 3 of \cite{muraveva2015new}.
 A glance at the  coefficients listed in this table and the   recent studies reveals the following features of the $Ks$- band PLZ relation of RRLs:  the period 
slopes
vary 
 from -2.73  \citep{muraveva2015new} to -2.101 \citep{bono2003pulsational}; the metallicity slopes 
vary from 0.03 $\pm$ 0.07 \citep{muraveva2015new} to 0.231$\pm$0.012\citep{bono2003pulsational};  the zero points 
vary from from $-1.17\pm0.1  $  \citep{clementini2017gaia} to $-0.769\pm0.088$ \citep{dambis2013rr}, and  all
these coefficients depend on the metalicity scale and the zero point calibration  method.
 
Our adopted period slope for Ks-band PLZ relation is consistent (within $1\sigma$ of uncertainties) with all previous empirical studies that uses Milky Way RRL samples and with \cite{catelan2004rr} theoretical study.  It also  (within $2\sigma$ of uncertainties) with the previous empirical studies that 
uses LMC RRL sample \citep[e.g.][]{borissova2009properties,muraveva2015new}, but it is slightly shallower than the previous theoretical  studies \citep[e.g.][]{bono2003pulsational,marconi2015new}. We   obtained a  significant  metallicity dependence of absolute magnitudes in the $Ks$-band like  theoretical studies \citep[e.g.][]{bono2003pulsational,catelan2004rr,marconi2015new}   and the recent empirical studies \citep[e.g.][]{muraveva2018rr,neeley2019standard,braga2018rr} which was not the case in most of the previous empirical studies.  
The zero point of our calibrated Ks-band PLZ relation 
at $[Fe/H]=-1.5$ and period of P=0.5238d  ($M_{Ks}=-0.30\pm0.08$) is only slightly ($ \sim$ 0.07 mag) fainter than the resent 
\cite{muraveva2018rr} calibration ($M_{Ks}=-0.37\pm0.11$).

\subsubsection{Comparison with Gaia DR2 parallax (Gaia DR2 zero-point offset)}
 A number of studies \cite[eg.][]{brown2018gaia,arenou2018gaia,luri2018gaia,stassun2018evidence,
riess2018milky,zinn2018confirmation}  confirmed that \textit{Gaia} DR2 parallaxes have a systematic zero-point error, which depends 
on the type of object and on position in the sky.  As an additional sanity check of our PLZ and LZ relations, we compared \textit{Gaia} DR2 trigonometric parallaxes with photometric parallaxes inferred from our calibrated PLZ relation to estimate the zero-point offset for \textit{Gaia} DR2 parallaxes.
For the comparison, we selected 846 sample RRLs that have reliable trigonometric parallaxes from \textit{Gaia} DR2 and then  used our calibrated W1-band PLZ relation (equation~\ref{eq:1}) to estimate their photometric 
parallaxes. 

\begin{figure}
\includegraphics[width=\columnwidth]{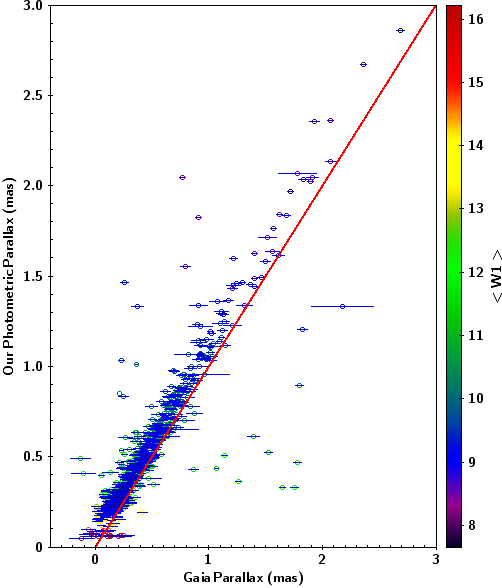}
    \caption{Comparison of the photometric parallaxes inferred from our calibrated $W1$-band PLZ relation with the corresponding DR2 parallaxes for 846 RRLs which have reliable estimates in both. The solid lines represent the bisectors and the auxiliary axis represent the W1-band mean magnitude.}
    \label{fig:parallax_off}
\end{figure}

Fig.\ref{fig:parallax_off}  compares our photometric parallaxes inferred from our calibrated W1-band PLZ relation and \textit{Gaia} DR2 parallaxes for 846 RRLs. As expected,  there is an overall systematic difference between the \textit{Gaia} DR2 parallaxes and our photometric parallaxes which depends on magnitude. The non-weighted mean difference between \textit{Gaia} DR2 and our photometric parallaxes is equal to  $-0.065\pm0.006$ mas. 
For comparison, in their validation of \textit{Gaia} DR2 parallaxes for
RR Lyrae stars, \cite{arenou2018gaia}  find zero-point offsets of $ -0.056 \pm 0.005$  mas for RRLs in the
\textit{Gaia} DR2 catalogue and $-0.033 \pm 0.009  $ mas  for RRLs in the GCVS.

Though, our estimated zero-point offset for \textit{Gaia} DR2 parallax is slightly larger than the published estimates mentioned above,  it agrees marginally, and this shows that \textit{Gaia} DR2 astrometry has  dramatically narrowed the gap between stasticall parallax and  trigonometric parallax methods to find distance and absolute magnitude of RRLs that have been existing in the past studies before \textit{Gaia} DR2.

\subsection{Distance Measurements}
In this section, we measure distances to the Galactic Center and LMC to check the performance of our calibrated PLZ and LZ relations.
\subsubsection{Distance to the Galactic Center}
 Distance to the Galactic center ($R_{\odot}$) is important for the studies on the Galactic structure and dynamics. So far many authors have determined $R_{\odot}$ using different methods and different distance indicators. Here, we improve $R_{\odot}$ estimated by three authors using  a similar procedure as \cite{dambis2013rr}. Thus, we use our calibrated PLZ or LZ relations (equation \ref{eq:1}-\ref{eq:5}) for their Galactic bulge RRLs sample by taking $ <[Fe/H]> = -1.0 $  as an average metallicity for Galactic bulge RRLs \citep{walker1991metallicity} to rescale their $R_{\odot}$.  Then, our best $R_{\odot}$ value is determined from the non weighted average of these three rescaled $R_{\odot}$.
 
The solar Galactocentric distance,
$8.3\pm1$ kpc, was inferred by \cite{carney1995distance} using their $K$- band observations on the California Institute of Technology (CIT) system ($K_{\text{CIT}}$)   for 58 RRLs in the Baade's window and by adopting   \cite{jones1992baade} PL relations. Hence, in order to rescale this $R_{\odot}$ estimate using our PLZ relation in Ks-band (eq.~\ref{eq:5}) we first transformed our PLZ relation in Ks-band into $K_{\text{CIT}}$-band by ignoring the negligible and statistically insignificant coefficient of the colour term in the transformation equation 
\footnote{\url{http://www.astro.caltech.edu/~jmc/2mass/v3/transformations/} }:
\begin{equation}
\begin{array}{rcl}
\langle M_{K,CIT}\rangle &=& -0.754(\pm0.08)+0.082(\pm0.026)\\
 & &[Fe/H]_{ZW}-2.33)\log P_F \label{eq:6}
\end{array}
\end{equation}

It is noticeable that the zero-point of our $K_{\text{CIT}}$-band  relation (eq.~\ref{eq:6}) in the Baade's window  
becomes   exactly equal to 
that of \cite{jones1992baade} used by the above authors and hence we can leave the estimate obtained by \cite{carney1995distance} unchanged,   $R_{\odot}=(8.30\pm1.0)$ kpc. 

The $V$ band extinction corrected mean  magnitude of RRLs in the Baade's window, $<V_{0}>=15.44\pm0.05 $, was estimated by \cite{collinge2006catalog} from
 the Optical Gravitational Lensing Experiment (OGLE II) data. Using this result and our calibrated LZ relation (eq.~\ref{eq:3}) for the Baade's window, we  estimated the Galactic center distance module  $ \mu_{GC}=14.631 \pm0.128$ which corresponds
to $ R_{\odot}=8.44\pm0.49$~kpc.  
 
\cite{groenewegen2008distance} also presented a catalog of 37 RRLs in the Galactic bulge that contains the dereddened $Ks$-band magnitudes and periods from their observations. This compilation without five bright outliers together with our calibrated PLZ relation in $Ks$-band at $<[Fe/H]>=-1.0$ 
offered Galactic center distance modulus estimate  of  $14.29 \pm 0.1$~ corresponding to~$R_{\odot}$=7.21$\pm0.62$~ kpc. 

Finally, we adopt the unweighted mean of our three $R_{\odot}$ measurement,   $R_{\odot}=7.99\pm0.48$, as our best $R_{\odot}$ estimate.  This result is consistent  with the  result of \cite{abuter2019geometric} ($ R_{\odot}=8178 \pm 13_{stat.} \pm 22_{sys.} pc$), which is estimated based on the orbital parameters of S2 star moving around the central massive black hole Sgr A$^{\ast}$;  but, it is less precise and shorter than the recent ones.   
 
Our result for $R_{\odot}$ combined with the  \cite{reid2004proper}  proper motion for the Sgr A$^{*}$ along Galactic longitude( $\mu_{SgrA^{*}} $=6.379$\pm$0.026 mas yr$^{-1}$) implies a solar velocity of  $ -242\pm15$\,kms$ ^{-1}$.  This is   slightly smaller than the recent \cite{hayes2018constraining}  estimate ($ -253\pm6$\,kms$ ^{-1}$) 
using \textit{Gaia} astrometry of Sagittarius Stream,  but it agrees marginally. The estimated solar velocity is also  consistent with the one ($\sim$ 238.1 km/s) that can be derived using the Oort constants calculated by \cite{li2019galactic} and the solar peculiar motion of \cite{ding2019local}.  This and our kinematic result in Table \ref{tab:table_3} lead us to conclude that both the inner halo and 
thick disk RRLs in the solar neighborhood undergoes a prograde rotation with a velocity of  23$\pm$17\,kms$ ^{-1}$
and 196$\pm$16\,kms$ ^{-1}$, respectively.

\subsubsection{The Distance Modulus of the LMC}
Another common distance measurement that can test our calibrated PLZ and LZ  relations  is the distance modulus of the LMC, whose value has been measured in countless studies using
different distance indicators and independent techniques 
 \citep[see e.g.][for compilation of the values in the literature]{gibson1999distance,benedict2002astrometry,
 clementini2003distance,de2014clustering,muraveva2018rr}.

In this paper, we measured the LMC distance modulus by applying our calibrated $Ks$-band PLZ relation (\ref{eq:5})   to a sample of 70 LMC RRLs compiled by \cite{muraveva2015new}. For these 70 RRLs \cite{muraveva2015new} took the pulsation mode and periods from the OGLE III catalog \citep{soszynski2009optical} and photometry
data from the VMC survey \citep{cioni2011vmc} to estimate the $Ks$-band intensity-averaged
magnitudes, and spectroscopic metallicities from  \cite{gratton2004metal}, which are 0.06 dex more metal-rich than  the \cite{zinn1984globular} 
scale. We used our calibrated $Ks$ band PLZ relation to determine the distance moduli for each of 70 LMC RRLs. We then computed the unweighted mean of the 
 individual distance modulus estimates as the distance module of LMC, which is found to be   18.46$\pm$0.09. Our LMC distance 
modulus estimate agrees well with the most precise LMC distance module $18.477 \pm 0.004_{(stat)} \pm 0.026_{(syst)} $ inferred by \cite{pietrzynski2019distance} based on eclipsing binaries.

\section{Conclusions}
\label{sec:Conclusions}
In this paper we compiled homogenized photometric (V-,Ks- and W1-band magnitudes), spectroscopic (radial velocity and metallicity) and astrometric
(proper motions and parallaxes) data for a sample of 850 Galactic field  RR Lyrae type variables. The compilation includes our new spectroscopic (for 448 RRLs) and photometric (for 251 RRLs ) observation data in addition to 
our previous  compilation\citep{dambis2013rr} (for 402 RRLs). 
The astrometric data for all RRL samples are obtained from \textit{Gaia} DR2. We used photometric data for a subsample of RRLs and \cite{drimmel2003three} 3D interstellar extinction model to calibrate the $(V-Ks)_0$ and $(V-W1)_0$ intrinsic colours of RRLs in terms of period and metallicity. Bimodal stastical parallax solutions for 
 a subsample of RRLs with close-to-solar Galactocentric distances yields the zero point correction for our  adopted LZ and PLZ relations, 
and  the kinematical parameters for the halo and thick-disc  RRL populations. 

Our  kinematical results indicate that the mean rotational velocity of the halo and disk RRLs in the solar neighborhood 
are  23$\pm$17\,kms$ ^{-1}$ and 196$\pm$16\,kms$ ^{-1}$, respectively, with respect to inertial  Galactic center reference. Moreover, the velocity dispersions for both halo and disk RRL population are non-isotropic $(\sigma V_{R}>\sigma V_{\phi}>\sigma V_{\theta})$ in the
solar neighborhood.
Our kinematical results are in good agreement with previous  statistical-parallax studies, but are considerably more accurate. Our currently largest homogeneous sample allowed us to reduce the uncertainty of RRLs distances down to $\sim +4.5\%$.  Thus, our kinematic results represent the current best estimates of this kind of studies
 and are at the same time quite consistent
with those obtained using other techniques. 

Our calibrated LZ and PLZ relations also agree well with those reported in published studies
 based on statistical parallax analysis. However, the zero points of our calibrated relations are fainter ($\sim$0.1 mag) than  other published estimates based on different methods. The reason for this is mostly due to underlying systematics in our analysis. Using our LZ and PLZ relations, we measured the solar Galactocentric distance   ($R_0=7.99\pm0.48$\,kpc) and the LMC distance
modulus  (DM LMC = 18.46$\pm$0.09).  Our $R_0$ estimate is  in good agreement with the 
recent results using eclipsing binaries  and  our LMC distance modulus agrees within $\sim$0.04 mag 
with the current precise 
measurement of \cite{pietrzynski2019distance}. From the comparison of \textit{Gaia} DR2 parallaxes with our photometric parallaxes that are estimated using our W1-band PLZ relation, we infer $-0.065\pm0.006$  \textit{Gaia} DR2 parallax offset. The magnitude of this offset is slightly larger than the published values. 

Using our largest RRL database, we will  investigate the structure and history
of our Galaxy in near future by combining 
abundance analysis with kinematics.  Further more, we will calibrate LZ and PLZ relations of RRLs using Bayesian Method. In the nearest future we expect to obtain more refined results concerning the kinematics and distance scale of Galactic RR Lyraes by expanding considerably the sample of these variables with 6D phase-space data provided by the recently published and forthcoming astrometric, photometric, and spectroscopic surveys and releases. These include primarily the next Gaia data releases (the third Gaia early data release with considerably improved per motions is expected by the end of 2020), PanStarrs and Zwicky Transient Factory optical photometric surveys, the available and forthcoming products based on the WISE mid-infrared space mission (unwise, catWISE, etc.), as well as extensive SDSS and LAMOST spectroscopic data.

\section*{Acknowledgements}
We thank anonymous reviewers for their valuable comments, which improved the final version of the article. All new spectroscopic observations reported in this paper were obtained with the Southern
African Large Telescope (SALT)
under programs 2015-2-SCI-043 and 2016-1-MLT-003 (PI: Alexei Kniazev).
A.\,Y.\,K. acknowledges support from the National Research Foundation (NRF) of South
Africa. T.\,D.\,M. acknowledges support from  Ethiopian Space Science and Technology Institute (ESSTI) and Bahir Dar University. This work
was supported by the Russian Foundation for Basic
Research (grants Nos. 18-02-00890 and 19-02-00611).
This publication makes use of data products from: the Wide-field Infrared Survey Explorer, which is a joint project of the University of California, Los Angeles, and the Jet Propulsion Laboratory/California Institute of Technology, and NEOWISE, which is a project of the Jet Propulsion Laboratory/California Institute of Technology. WISE and NEOWISE are funded by the National Aeronautics and Space Administration; the European Space Agency (ESA) mission
{\it Gaia} (\url{https://www.cosmos.esa.int/gaia}), processed by the {\it Gaia}
Data Processing and Analysis Consortium (DPAC,
\url{https://www.cosmos.esa.int/web/gaia/dpac/consortium}). Funding for the DPAC
has been provided by national institutions, in particular the institutions
participating in the {\it Gaia} Multilateral Agreement;the Two-Micron All-Sky
Survey, which is a joint project of the University of Massachusetts
and the Infrared Processing and Analysis Center/California
Institute of Technology, funded by the National Aeronautics and
Space Administration and the National Science Foundation.


\bibliographystyle{mnras}
\bibliography{tes1} 


\appendix

\section{Some extra material}

Additional Supporting Information may be found in the online version of this article:
Table S1. The catalogue of observational data for 850 Galactic
field RR Lyrae variables


\bsp	
\label{lastpage}
\end{document}